\documentclass{article}
\usepackage{epsfig,amsmath,amssymb,hyperref,color}
\usepackage[parfill]{parskip}

\def\Lobad{{{\Bbb H}^{d-1}}}
\def\Re{{ \rm Re}\,}
\def\Im{{ \rm Im}\,}
\def\Arg {{ \rm Arg}}

\def\x{{\mathrm x}}
\def\P{{\bf P}}
\def\Q{{\bf Q}}

\def\Xd{dS_d}
\def\Futcone{{\Gamma^+}}
\def\d{{\partial}}
\def\ndir{{ \rm n}}
\def\Xisign{\epsilon}
\def\Xiang{a}
\def\aa{{\mathrm a}}

\def\FS{ F}
\def\CG{{ C}}
\def\Ylm{{ Y}_{l,M}}
\def\Yslm{{ Y}^*_{l,M}}
\def\Nq{n}
\def\PR{\Pi}
\def\DEL{\hat\delta}
\def\Xdc{dS^{(c)}_d}
\def\cnew{\gamma}
\def\O{ \rm O}
\def\warp{a}

\begin{document}

\title{Quantum theory on Lobatchevski spaces}
\author{Ugo Moschella$^1$ and Richard Schaeffer$^2$ \\$^1$Universit\`a dell'Insubria, 22100 Como, Italia, and INFN, Sez. di Milano,
Italia\\$^2$Service de Physique Th\'eorique, CEA -
Saclay, Gif-sur-Yvette, France}

\maketitle

\begin{abstract}{In this paper we set up a general formalism to
deal with quantum theories on a Lobatchevski space, i.e. a
spatial manifold that is homogeneous, isotropic and has
negative curvature. The heart of our approach  is the
construction of a suitable basis of plane waves which are
eigenfunctions of the Laplace-Beltrami operator relative to the
geometry of the curved space. These functions were previously
introduced in the mathematical literature in the context of
group theory; here we revisit and adapt the formalism in a way
specific for quantum mechanics. Our developments render dealing
with {{Lobatchevski spaces}}, which used to be quite difficult
and source of controversies, easily tractable. Applications to
the Milne and de Sitter universes are discussed as examples.}
\end{abstract}

\section{Introduction.}
There are several good reasons to study quantum theories on a
Lobatchevski space. The first reason is simply to extend our
knowledge and skills in quantization. The negative curvature of
this model together with its non compactness may, and indeed
do, give rise to new and unforeseen phenomena which ordinary
(canonical) quantization procedures on curved spacetimes
\cite{Birrell} are not prepared to deal with.

Secondly, it has for long been advocated by Callan and Wilczek
\cite{Callan} that a negative curvature may act as a covariant
regularizer of the infrared problem in a better way than
putting a quantum system in a box or on a sphere. An intuitive
way to understand this viewpoint is as follows: since in this
geometry the volume of a sphere increases exponentially with
its radius, the flux created by a central charge decreases
accordingly and photons behave as if they have a "mass". Many
quantum field models have indeed been studied on this
background geometry, in particular the two-dimensional
Liouville model \cite{Zamolo} and a whole class of conformally
invariant models.

{From} the viewpoint of astrophysics and  cosmology,
Friedmann-Robertson-Walker open  models (more properly, models
with negative spatial curvature),  {}{first introduced in the
inflationary context in the early eighties \cite{Gott}, became
rather popular
\cite{Ratra94,RatraPeebles,kamionkowski,Bucher:1994gb}}  in the
mid-nineties, when the belief was that a negative curvature of
the space could explain the mass content and the expansion rate
of the universe. At that time it was realized that in some
cases \cite{Sasaki} there were troubles with canonical
quantization in such a geometry, although with adaptation the
latter can be used to recover the standard
\cite{Gibbons,BunchDavies,bgm} Klein-Gordon quantum field in
the special case of the open de Sitter manifold and some
indications exist \cite{MS} for possible ways out in more
general situations. One difficulty  was the appearance of
supercurvature modes for theories of sufficiently low mass.
Those modes are not square integrable on the Lobatchevski
manifold itself (even not in generalized sense as the
exponentials of the flat case do) and therefore do not fit in
the standard formalism of quantum mechanics. Even though the
physical meaning of those supercurvature modes is yet unclear,
they are not expected to provide a sizeable contribution to the
microwave background fluctuations \cite{Gorski}. They might
however be relevant in other domains of physics $\ldots$ and
even in cosmology that from time to time (and even often) gives
rise to surprises.

The interest in open inflationary models subsequently dropped
with the measurements of the fluctuations of the Cosmic
Microwave Background \cite{BMRG} indicating that the universe
is most likely flat. However, with the spectacular and
unprecedented precision of the forthcoming measurements
scheduled by the PLANCK satellite \cite{PLANCK}, any small
deviation from flatness will have to be mastered securely. This
calls for a revision and a complete solution of the problem of
describing quantum fluctuations in a negatively curved space
(while the positively curved case is completely under control).

{From} the mathematical viewpoint, part of the material we are
going to present belongs to the chapter of harmonic analysis on
symmetric spaces (see e.g. \cite{helg}; \cite{prog} gives an
up-to-date account of the topics). Of course there exists a
vast literature already on the more specific subject of
Lobatchevski spaces. These mathematical results however are
often formulated in a rather abstract way. Also, the
square-integrability hypothesis that is a pillar of harmonic
analysis may be violated in physically interesting situations
and should only be considered as a starting point. On the other
hand, most if not all approaches in the physical literature
involve series expansions in terms of special functions which
hide to a large extent the underlying symmetries and ultimately
the physics.

We have therefore decided to reconsider the subject from the
beginning and we have found some new ways to handle
efficiently, and rather simply, the technical difficulties that
arise from the lack of a commutative group of space
translations. Our approach is especially aimed for physics; it
remains practical and accessible and has also the advantage to
avoid the use of theorems concerning expansions in bases of
special functions (those expansions actually result from our
approach). Another valuable point is that we work solely within
the physical space-time and do not rely on any extension or
completion of the latter to regions that are not covered by the
open chart, as is the case in previous studies. Our scheme is
sufficiently flexible to allow the study of the general
dimension in a single step.

We focus on the study of a basis for the the standard Hilbert
space of square-integrable functions on a $(d-1)$-dimensional
Lobatchevski space (so that the spacetime has $d$ dimensions).
This Hilbert space is not expected to be sufficient
\cite{Sasaki,MS} to describe all the interesting physical
quantum theories in case of negative spatial curvature. Modes
that are not square-integrable, however, warrant a separate
specific study, and will be considered elsewhere
\cite{MSfollow} along the same line of thought.

In Section   \ref{sect:coord} we give an introductory
presentation of the geometry of the Lobatchevski space and of
the "absolute" \cite{gelfand} of that space which can be
identified with the space of momentum directions. We give also
a quite detailed description of the coordinate systems we are
going to use.

In Section   \ref{sect:laplacian} we display  the unnormalized
eigenmodes of the Laplace-Beltrami operator following the
approach described in \cite{gelfand}. These solutions are
"plane wavefunctions"  in the sense they have constant values
on hyperplanes. Strangely enough, these solutions have been
only rarely  used to study quantum theories in a Lobatchevski
space in the physical literature, {but they constitute the most
natural possibility}, and actually a cornerstone, to phrase
Lobatcehvskian quantum mechanics in strict analogy with
Euclidean quantum mechanics. An important technical point is
provided in Section \ref{sec:modereps} where we construct some
useful integral representations of the above eigenmodes; these
representations are linked to the parabolic coordinate system
introduced in Section   \ref{sect:coord} and are inspired by
our earlier work \cite{Bertola}. These representations allow to
trivialize $(d-2)$ integrations when a square-integrable
wavefunction is projected on a mode.  The relation of our modes
with the basis constructed in spherical coordinates (see e.g.
\cite{Ratra94} and references therein) is also discussed.

In Sections   \ref{sec:scalprod},   \ref{sec:inttrans} and
  \ref{sec:projq} we set up the basic ingredients for quantum
mechanics by building an orthonormal basis of modes; in
particular we show how to deal with the integral
representations by computing the $L^2$ scalar product of the
modes. This naturally leads to the introduction of a
Fourier-type transform  of an arbitrary $L^2$ function and to an
inversion formula obtained here by means of the
Kontorovich-Lebedev transform.

We conclude the paper by discussing two physical applications
to models of QFT that are playing a central role in
contemporary cosmology.

In Section   \ref{sec:milne} we revisit Milne QFT. There has been
a regain of interest in this  model in recent times coming from
the very different perspectives of string theory \cite{Khoury,
Zamolo} and of observational cosmology \cite{Kutschera}. Here
we give a clean treatment of Milne's QFT based on an expansion
of the Minkowskian exponential plane waves onto the basis of
Lobatchevski modes that we have been constructing.

In the following Section   \ref{sec:dSQFT}, we study QFT on the
open de Sitter universe at any dimension. This has proven to be
technically very difficult and has been source of
controversies, even for the quantification in $L^2$ space,
which is the aim of the present paper \cite{Sasaki,MS}. A
special attention is given to the two dimensional case, which
shares many features of the general case, but is extremely
simple.

The anti-de Sitter case is of obvious interest: the
Lobatchevski manifold is identical to the Euclidean anti-de
Sitter universe. The physics however is however very different
and therefore we have not included the AdS case in the present
paper.

\section{Coordinates. \label{sect:coord}}
In this section we describe the relevant geometrical setup for
our construction, following the ideas and the layout of
Gel'fand, Graev and Vilenkin \cite{gelfand}.

\begin{figure}[h]
\begin{center}
\includegraphics[height=8cm]{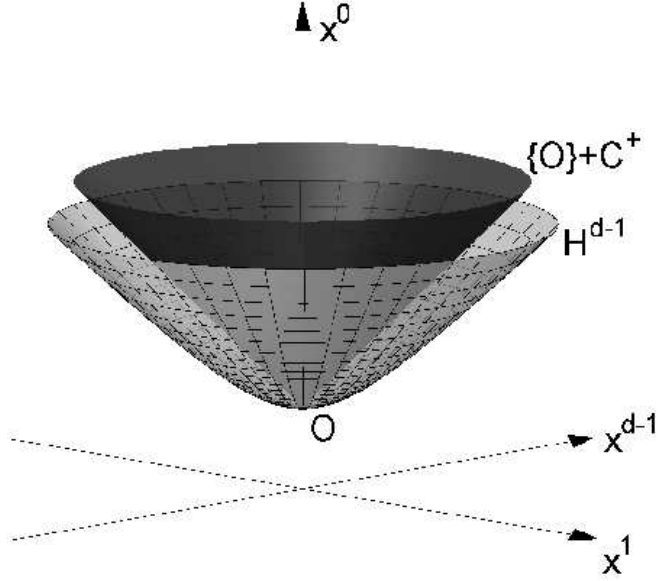}
\caption{\em A space of constant negative curvature embedded in an
ambient Minkowski spacetime with one dimension more. The manifold
$\Lobad$ does not intersect the lightcone issued at any of its
points (the origin $\O$ in the figure); this shows pictorially that
the surface is a spacelike manifold i.e. a Riemannian model of
space.} \label{fig:sigma}
\end{center}
\end{figure}

Let ${\mathbb M}^d$ be a $d$-dimensional Minkowski spacetime;
an event $x$ has inertial coordinates $x^0,\ldots x^{d-1}$ and
the scalar product of two such events is given by
\begin{equation}
x\cdot {x'} = x^{0}{x'^{0}}-{x^{1}}{x'^{1}}-\ldots-
{x^{d-1}}x'^{d-1} = \eta_{\mu\nu}{x}^{\mu}x'^{\nu} .
\end{equation}
In ${\mathbb M}^d$ we consider the manifold (Fig.
  \ref{fig:sigma})
\begin{equation}
\Lobad  = \{x\in{\mathbb M}^d:\ x^2 = x\cdot x = 1,\;\; x^{0}>
0\}.
\end{equation}

The manifold $\Lobad$ models a $(d-1)$-dimensional Lobatchevski
hyperbolic space\footnote{We use here the convention mostly
adopted by topologists in using the notation $\Lobad$. The
index $d-1$ refers to the actual dimension of the manifold in
the same way as \( {\Bbb S}^{d-1} = \{x \in {\mathbb R}^d,\,\,
{x^1}^2 + \ldots {x^{d}}^2 = 1\} \) denotes a
$(d-1)$-dimensional sphere embedded in a space of dimension
$d$. Note however that the conventional notation for the
hypersurface of the sphere ${\Bbb S}^{d-1}$ is
$\omega_d={2\pi^{{d}/{2}}}/{\Gamma\left({d}/{2}  \right)}$.
 \label{foot:d-spheres}}
. It is a homogeneous space under the action of the restricted
Lorentz group of the ambient spacetime $SO_0(1,d-1)$. The
Riemannian metric $dl^2_{d-1}$ is obtained by restriction of
the ambient Lorentzian metric to $\Lobad$:
\begin{equation}
dl^2_{d-1} = \left. -\left\{ {\left(d\x^{0}  \right)}^2 -
{\left(d\x^{1}  \right)}^2-\ldots -
{\left(d\x^{d-1}  \right)}^2  \right\}  \right|_{\Lobad} .
\label{opmetr}
\end{equation}

\subsection{Parabolic coordinates.}\label{pcsub}
Many interesting coordinate systems on $\Lobad$  arise from
particular decompositions of the symmetry group $SO_0(1,d-1)$.
In this paper we will mainly use the following one:
\begin{equation}
x( r,\x) = \left\{\begin{array}{lclcl}
x^{0} &=& \frac{1+\x^2 + r^2}{2 r}   \\
x^{i} &=&   \frac{\x^i}{ r}\\
x^{{d-1}} &=&  \frac{1-\x^2 - r^2}{2 r}
\end{array}   \right.
\label{dscoor},\;\;\;\;0<r<\infty,\;\;\;\;\x \in {\mathbb R}^{d-2}.
\end{equation}
In these coordinates the metric and the invariant volume form
have the following explicit expressions:
\begin{eqnarray}
\ \ \ \ \ \ \ \ \ \ \ \ \ \ \ \ \ \ \ dl^2_{d-1} = \frac{d r^2+ d\x^2}{ r^{2}}, \quad \label{metr}
d\mu(x) = r^{-(d-2)} \, \frac{d r}{ r} \, d\x , \label{volform}
\end{eqnarray}
while the scalar product of two event is written
\begin{equation}
x( r,\x) \cdot x'( r',\x')= \frac{\left({\x}-{\x'}  \right)^2+ r^2
+{ r'}^2}{2 r r'}  . \label{xxsccoord}
\end{equation}
Equations (\ref{volform}) and (\ref{xxsccoord}) show that  the
measure and the scalar product  are invariant w.r.t.
translations in the $\x$ coordinates and this explain why this
system is so important and useful. Dirac's delta distribution
on the hyperboloid $\Lobad$ is understood w.r.t. the invariant
measure
\begin{equation}
\int_{\Lobad} d\mu(x')\   \delta(x,x') f(x')  = f(x);
\end{equation}
specifically, in the previous coordinate system it is written
as follows: \begin{equation} \delta(x,x') =
 r^{{d-1}}\delta( r-  r')\delta(\x-\x'). \end{equation}

\begin{figure}[h]
\begin{center}
\includegraphics[height=8cm]{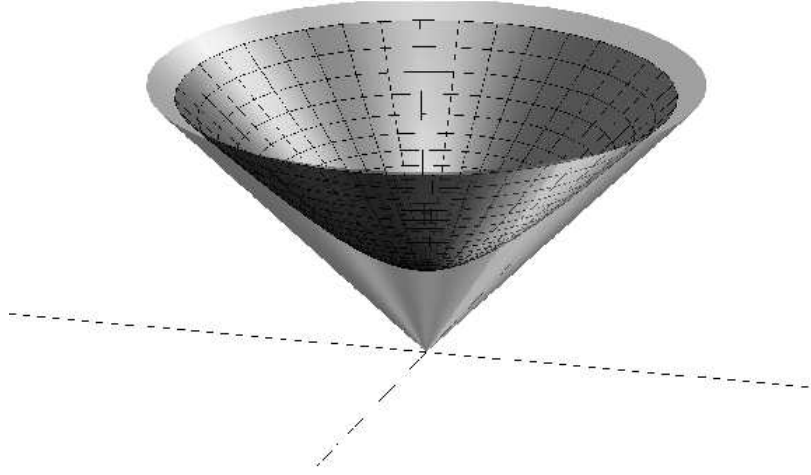}
\caption{\em A sight of the asymptotic future lightcone of the ambient
spacetime; the vectors $\xi$  belonging to $C^+$ play the role of
momentum directions.} \label{fig:absolute}
\end{center}
\end{figure}

The future light-cone of the ambient spacetime (Fig.
  \ref{fig:absolute}) plays a crucial role in our construction:
\begin{equation}
C^{+}= \{ \xi^2 = \xi\cdot \xi = 0,\;\;\; \xi^{0}> 0\}.
\end{equation}
A useful parametrization for vectors of $C^{+}$ corresponding
to (\ref{dscoor}) is the following:
\begin{equation}
 \xi(\lambda,\eta)=\left\{\begin{array}{lcc}
\xi^{0} &=&  \frac\lambda 2 (1+ {\eta}^2)    \\
\xi^{i} &=&  \lambda {\eta^i}\\
\xi^{{d-1}} &=& \frac\lambda 2 (1- {\eta}^2)
\end{array}   \right.
\label{abscoor} \;\;\;\;0<\lambda<\infty,\;\;\;\;\eta \in {\mathbb
R}^{d-2}.
\end{equation}

\begin{figure}[h]
\begin{center}
\includegraphics[height=3.5cm]{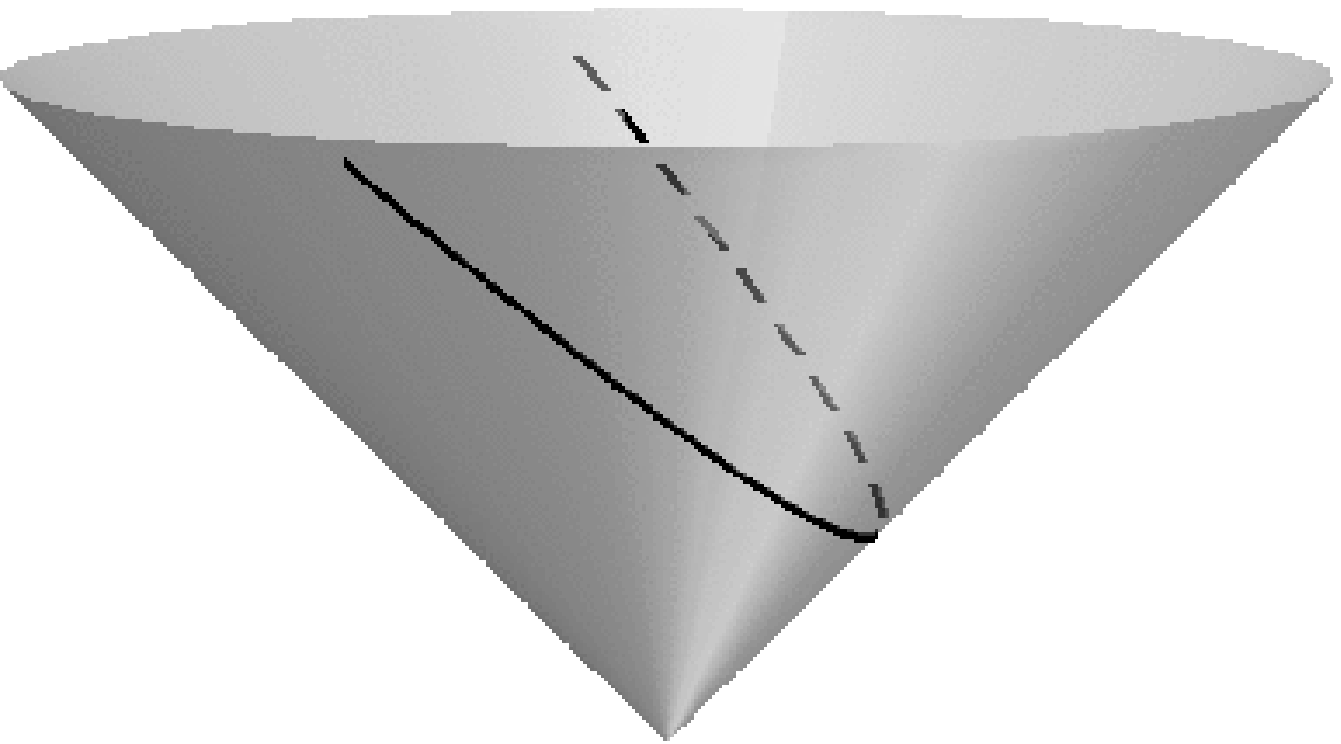}
\ \ \ \ \ \ \ \ \ \
\includegraphics[height=3.5cm]{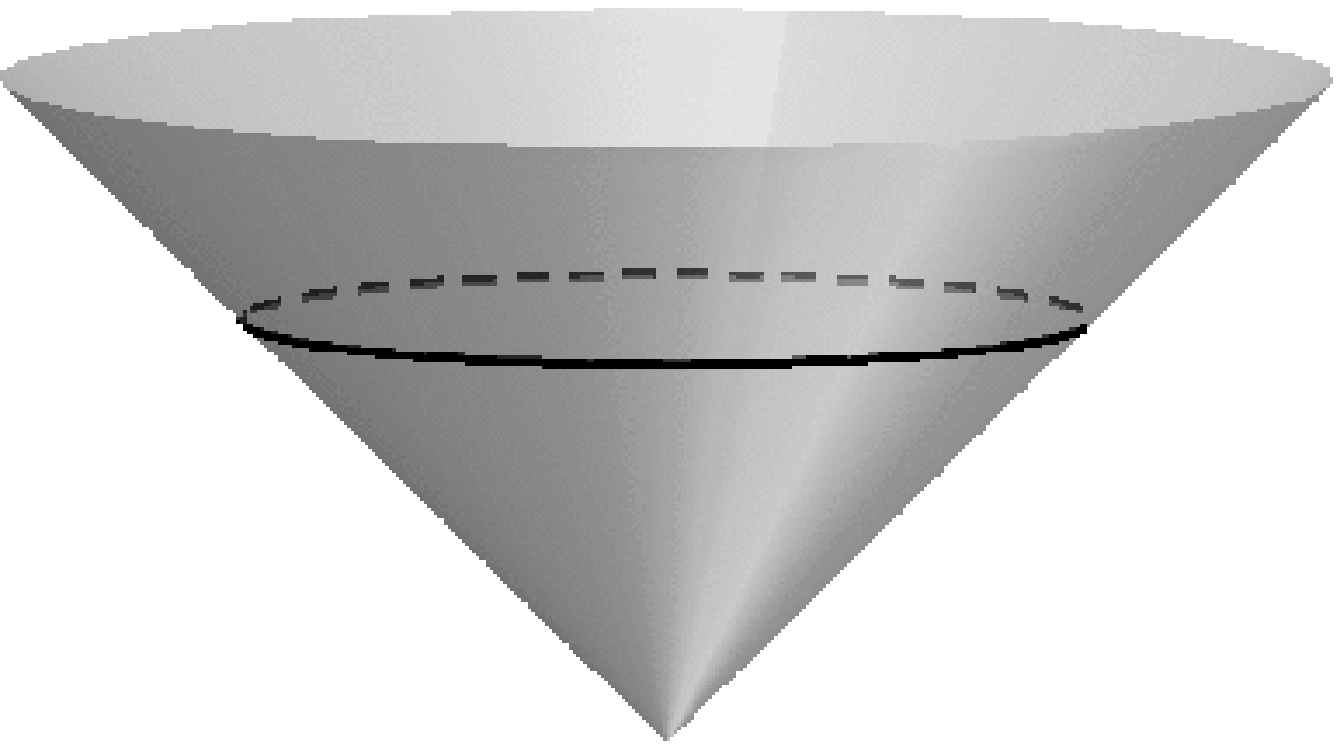}\end{center}
\caption{\em A view of the parabolic and spherical bases of the
absolute. \label{fig:absolutepar}}
\end{figure}

From both the mathematical and the physical viewpoints what
matters is \cite{gelfand} the "absolute" of the Lobatchevski
space $\Lobad$: this is the set of linear generators of the
future light-cone, i.e. the light-cone modulo dilatations. The
previous parametrization gives in particular a parabolic
parametrization of the absolute that is visualized (Fig.
  \ref{fig:absolutepar})   as the parabolic section $\lambda = 1$
of $C^+$:
\begin{equation}
\xi(\eta) \equiv \xi(1,\eta)=\left\{\begin{array}{lcc}
\xi^{0} &=&  \frac12 (1+ {\eta}^2)    \\
\xi^{i} &=&  {\eta^i}\\
\xi^{{d-1}} &=& \frac12 (1- {\eta}^2)
\end{array}   \right. \;\;\;\;\eta \in {\mathbb R}^{d-2}.
\label{conecoor}
\end{equation}
Correspondingly, the measure on the absolute  is normalized as
follows:
 \begin{equation}
d\mu(\xi) = d\eta. \end{equation}

\begin{figure}[h]
\begin{center}
\includegraphics[height=5cm]{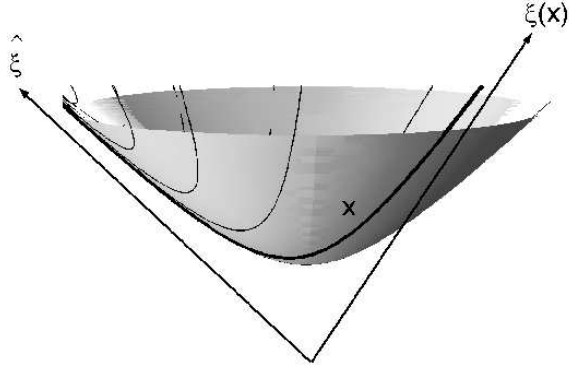}\end{center}
\caption{\em The choice of the vector $\hat{\xi}$,given the point $x$, uniquely determines the
vector $\xi(\x)$. The latter is the vector obtained by
intersecting the cone with the plane containing the vector
$\hat{\xi}$ and the point $x$ (this plane actually contains the
whole curve $x(r,\x),\ r>0)$. "Infinity" on the curve $x(r,\x)$ is
at $ r =0$ in the direction $\xi(\x)$ -which depend on the chosen
point $\x$- and at $ r=\infty$ in the direction $\hat{\xi}$.
\label{fig:xparam}}
\end{figure}

Dirac's delta on the cone will actually mean Dirac's delta on
the absolute:
 \begin{equation} \delta(\xi,\xi') =
\delta(\eta-\eta'). \end{equation} By using the coordinate systems
(\ref{dscoor}) and  (\ref{conecoor}) the scalar products ${x
\cdot \xi}$  and $\xi\cdot \xi'$ are written as follows:
\begin{eqnarray}
{x( r,\x) \cdot \xi(\eta)} &=&
\frac{\left({\x}-{\eta}  \right)^2}{2 r} + \frac{ r}{2} =
\frac{\xi(\x) \cdot \xi(\eta)}{ r} + \frac{ r}{2} \label{xxicoord}\\
{\xi(\eta) \cdot \xi'(\eta')} &=&
\frac{\left({\eta}-{\eta'}  \right)^2}{2}
 \label{xixicoord}
\end{eqnarray}
where we have introduced the lightlike vector $\xi(\x)$
corresponding to $x(r,\x)$. Note that this correspondence is
tight to the choice of coordinates (\ref{dscoor}). Starting
from $\xi(\x)$ we may recover the point $x$ as follows (Fig.
  \ref{fig:xparam}):
\begin{equation} x(r,x) = \frac {\xi(\x)}{ r} +  r \hat\xi \label{xxi}\end{equation}
where $\hat{\xi} = \xi(0)=\left(\frac 12, 0 ,\ldots ,-\frac
12  \right)$ generates the only lightlike direction that the
parametrization (\ref{conecoor}) does not cover; this direction
is actually attained in the limit $ r\to \infty$ as shown by
Eq. (\ref{xxi}); one has that $\hat{\xi}\cdot x(r,\x) =
{1}/{2r} $.

%

\subsection{Spherical coordinates.}\label{scsub}
There is the other natural and widely used coordinate system on
the manifold $\Lobad$ where the rotation symmetry $SO(d-1)$ is
apparent:
\begin{eqnarray}
&&x(\psi,\phi) = \left\{
\begin{array}{l}
x^0 = \cosh \psi\\
x^1 = \sinh \psi \cos \phi^1 \\
\vdots \\
x^{d-2} =\sinh \psi \sin \phi^1 \ldots \sin\phi^{d-3}\cos \phi^{d-2}\\
x^{d-1} = \sinh \psi \sin \phi^1 \ldots \sin\phi^{d-3}\sin \phi^{d-2}
\end{array}  \right.
\cr\cr &&
\psi\in{\Bbb R},\;\;\; \phi = (\phi^1, ... , \phi^{d-2}),  \;\;\; 0<\phi^i<\pi,\;\;\;0<\phi^{d-2}<2\pi
.\label{sphericds}
\end{eqnarray}
Correspondingly, there is the important  (compact) spherical
basis $\xi^0=1$ for the absolute (see Fig.
(\ref{fig:absolutepar})):
\begin{eqnarray}&&\xi(\theta) =
\left\{
\begin{array}{l}
\xi^0= 1\\
\vec \xi = \ndir_\theta
\end{array}  \right.
= \left\{
\begin{array}{l}
\xi^0 = 1\\
\xi^1 = \cos \theta^1 \\
\vdots \\
\xi^{d-2} = \sin \theta^1 \ldots \sin\theta^{d-3}\cos \theta^{d-2}\\
\xi^{d-1} = \sin \theta^1 \ldots \sin\theta^{d-3}\sin \theta^{d-2}
\end{array}  \right.
\label{sphericcone}
\end{eqnarray}
with $ \theta = (\theta^1, ... , \theta^{d-2})$ and where $
\ndir_\theta$ is a unit vector ($  \ndir_\theta\cdot
\ndir_\theta=1$) pointing in  the direction identified by the
angles $\theta$. In these coordinates the scalar products are
written as follows
\begin{eqnarray}
x(\psi,\phi) \cdot x(\psi',\phi') &=& \cosh \psi \cosh \psi' - \sinh \psi\sinh \psi' \ { \ndir}_\phi \cdot { \ndir_{\phi'}}\\
{x(\psi,\phi) \cdot \xi(\theta)} &=& \cosh \psi - \sinh \psi \ { \ndir_\phi}\cdot { \ndir_\theta}\label{opq}\\
\xi(\theta) \cdot \xi(\theta') &=& 1 -  { \ndir_\theta}\cdot { \ndir_{\theta'}}
\end{eqnarray}
The measure on the absolute is the rotation invariant measure
normalised as follows:
\[ d\mu(\theta) =
\prod_{i=1}^{d-2}  (\sin\theta^i)^{d-2-i}   \,  d\theta^i \]

\section{Modes of the Laplacian. Plane waves.\label{sect:laplacian}}
Let us consider  the vector space ${\cal S}(\Lobad)$ of rapidly
decreasing complex functions defined on the manifold $\Lobad$
and let us introduce the natural scalar product w.r.t. the
invariant measure $d\mu(x)$:
\begin{equation}
\langle f,g \rangle = \int_{\Lobad}    {f}{^*}(x) g(x)
d\mu(x)\ . \label{hilbert}
\end{equation}
The space  ${\cal S}(\Lobad)$ can be completed to construct the
Hilbert space ${\cal H}=L^2(\Lobad,d\mu)$. By analogy with the
flat case, $\cal H$ is the natural Hilbert space one would
consider to study quantum mechanics on the homogeneous and
isotropic hyperbolic space $\Lobad$. To pursue this analogy,
the first object to be examined is the free Hamiltonian
operator
\begin{equation}
H = -\Delta\  , \end{equation} where $\Delta$ denotes the
Laplace-Beltrami operator  associated with the geometry
(\ref{opmetr}). The operator $-\Delta$ is self-adjoint on a
suitable domain of the Hilbert space $\cal H$ and its spectrum
is the set $[\frac 14 \left(d-2  \right)^2,\infty)$. The
eigenfunctions of the operator $-\Delta$ can be labeled by a
forward lightlike vector ${\xi} \in C^+$   and a real number
$q$ as follows \cite{gelfand}:
\begin{equation}
\psi_{iq} (x,{\xi} ) = \mbox{\em const} \left(x\cdot \xi
  \right)^{-\frac{d-2}{2} + iq }, \label{Sigwaves}
\end{equation}
and one easily verifies that
\footnote{This can be shown by using a specific coordinate
system,  for instance (\ref{dscoor}), or either by using the
inertial coordinates of the ambient spacetime, by introducing
the projection operator $h$ and the tangential derivative $D$
as follows:
\begin{equation}\label{}
h^{ \mu  \nu } = \eta^{ \mu  \nu } - {x^{\mu} x^{\nu}}, \ \ \ \ D^{
\mu } = h^{ \mu  \nu }{ \d}_{ \nu }= { \d}^{ \mu } -  \, x^{\mu}\,
x\cdot \d.
\end{equation}
For any function that is smooth in a neighborhood of the
manifold $\Lobad$ one has that
\begin{equation}  D^{ \mu } D_{ \mu } f = \Box f -
({d-2}) \ x\cdot \d f - \, x\cdot \d \ (x\cdot \d f),
\end{equation}
where $\Box$ is the wave operator in the $d$-dimensional
ambient spacetime. From this relation one can easily see that
\begin{equation}\label{}
D^{ \mu } D_{ \mu } \left( { x \cdot {\xi}}   \right)^{\alpha} =
\alpha(\alpha-1)\xi^2 \left( { x \cdot {\xi}}   \right)^{\alpha} -
{\alpha(\alpha+ d-2) } \left( { x \cdot {\xi}}   \right)^{\alpha}= -\
\alpha(\alpha + d - 2) \left( { x \cdot {\xi}}   \right)^{\alpha}
\end{equation}
Since $ -\Delta f = \left. D^\mu D_\mu \widehat f \
  \right|_{\Lobad},$ where $\widehat f$ is any smooth extension
of the function $f$ in a neighborhood of the manifold $\Lobad$
Eq. (\ref{eqlapl}) follows.}
\begin{equation}
-\Delta {\psi}_{iq} (x,{\xi} )= \left[\left({\frac{d-2}{2}}  \right)^2 +
q^2  \right]{\psi}_{iq } (x,{\xi}) = k^2 {\psi}_{iq } (x,{\xi}).
\label{eqlapl}
\end{equation}

For fixed $q$ and any $\lambda>0$, the vectors ${\xi}$ and
$\lambda{\xi}$ identify the same eigenfunction because of the
homogeneity properties of the expression  ({  \ref{Sigwaves}).
Therefore the modes (\ref{Sigwaves}) corresponding to real
values of the parameter $q$ and to vectors $\xi$ on the
absolute (i.e. on a basis of the asymptotic lightcone) can be
used to construct a basis of the Hilbert space $\cal H$ and are
the strict analogue of the purely imaginary exponentials $e^
{ip\cdot x}$ of the flat case. Indeed, like the exponentials,
also the wavefunctions (\ref{Sigwaves}) take  constant values
on planes, here the hyperplanes $x\cdot \xi = \mbox{const}$. In
this sense the modes ({  \ref{Sigwaves}) may be called ``plane
waves''.

As for the physical interpretation of the labels,  $k$ may be
thought of as the intensity  and $\xi$ as the ``direction'' of
a ``momentum'' vector identifying a mode. The analogy in flat
space would be parameterizing the plane wave by the modulus of
the momentum $p = (p\cdot p)^\frac12$ and by its direction
$\ndir$: $e^{ ip\, \ndir \cdot x}$.

This analogy goes one step further: we could have considered
another set of plane waves characterized by the opposite of the
modulus of the momentum $-p$ and by a direction $\ndir'$: $e^{-
ip\, \ndir' \cdot x}$. A trivial remark is that these modes
already belong to the set \{$e^ {ip \ndir \cdot x}$\} since
$\ndir $ can point in any direction.

We will see in the following section, that also the mode
$\left(x\cdot \xi   \right)^{-{\frac{d-2}{2}}- iq}$ can be expressed as a
superposition on the absolute (i.e. as an integral over $\xi'$
on a basis of the cone $C^+$) of the modes $\left(x\cdot \xi'
  \right)^{-{\frac{d-2}{2}} + iq }$. This means that to construct a basis
of the Hilbert space $\cal H$ we may restrict our attention to
the plane waves $\psi_{iq}(x,\xi)$, $q\geq 0$, $\xi$ on the
absolute.

We end this section by remarking that there are  purely
imaginary values of $q$ such that $k^2 \geq 0$, namely those
purely imaginary $q$ such that $|q| \leq \frac{d-1}2$. In
particular the mode corresponding to $k=0$ is constant in
space. These waves are not conventional in many respects; they
are real functions, do not oscillate and (superposition of
them) do not belong to the natural Hilbert space $\cal H$. This
means that a standard quantum mechanical interpretation is not
immediately available for them. Such cases have been considered
in the past \cite{Sasaki,MS} but their status is not
completely understood. We will discuss their possible role
elsewhere \cite{MSfollow}.

\section{Representations of the modes. \label{sec:modereps}}
There is no way to write the modes of the Laplace-Beltrami
operator that be more symmetric than the expression
(\ref{Sigwaves}): in that definition the modes appear as a
complex power of a quantity invariant under the action of the
symmetry group, exactly as it happens for the exponentials
$\exp(i{ p} \cdot x)$  in the flat case. The exponentials have
however the important property to be characters of the
translation group: this property is expressed by the relation
$\exp(ip\cdot x) \exp(ip \cdot y) = \exp(ip \cdot(x+y))$. The
lack of translation invariance of $\Lobad$ is a major technical
difficulty and, for our modes, there is nothing immediately
replacing this property of the exponentials. Therefore, from
the viewpoint of practical calculations, it is useful to
represent the modes (\ref{Sigwaves}) in terms of some integral
transform which be reminiscent of translational invariance.
Many representations are possible, in relation with different
choices of coordinates on $\Lobad$ and on the cone; we list
only the two that are more relevant for our purposes.

\subsection{Euler Representation.}
The following integral representations is an adaptation of the
Euler integral of the second kind expressing the Gamma
function; an interesting geometrical interpretation can be
based on the embedding of $\Lobad$ in the ambient Minkowski
spacetime ${\mathbb M}^d$:
\begin{equation}
{\left({x \cdot \xi}   \right) }^{- \frac{d-2}{2} + i  {q}}\, =
\frac{1}{{\Gamma}\left(\frac{d-2}{2} - i  {q}  \right)}\int
_{0}^{\infty}\frac{dR}{R}\   R^{\frac{d-2}{2} - i  {q}}\, e^{-R\,
{x\cdot\xi}} . \label{EuSigwaves}\end{equation} This representation
is valid if ${\Im}q> \Re\frac{d-2}{2} $ and ${\Re} (x\cdot\xi)
> 0.$  More generally one can perform the integration on the
complex plane as follows:
\begin{equation}
  {\left({x \cdot \xi}   \right) }^{- \frac{d-2}{2} + i  {q}}\, =
   \frac{{i^{-(\frac{d-2}{2} - i  {q})}}}{{\Gamma}\left(\frac{d-2}{2} - i  {q}  \right)}\int \frac{dR}{R}\
R^{\frac{d-2}{2} - i  {q}}\, e^{i R\,  {x\cdot\xi}}
 \label{comprep}
\end{equation}
where the integration contour in the complex $R$ plane is along
any half-line issued from the origin in the upper half-plane,
i.e. $0<\Arg (R)<\pi$.

\subsection{Fourier Representations.}
Another useful integral representations  can be obtained by
inserting at the RHS of Eq. (\ref{EuSigwaves}) the
representation (\ref{xxicoord}) of the scalar product
$\x\cdot\xi$ and Fourier transforming the Gaussian factor
appearing there:
\begin{eqnarray}
 \left({x\cdot\xi}  \right)^{-\frac{d-2}{2} + iq} &=&
 \frac{1}{\Gamma\left(\frac{d-2}{2}-i q  \right)}\left( \frac{ r}{2\pi}  \right)^{\frac{d-2}{2}}
\int \frac{d\kappa}{\kappa^{iq}}\ e^{-i\kappa\cdot\left({\x}-{\eta}  \right)}
\int_{0}^{\infty} \frac{dR}{R} \  \frac{ e^{-\frac 12  r{\kappa}
\left(\frac{1}{R} + R  \right)}}{R^{ i q}}
 \label{FSigwaves}
\cr
&=&
 \frac{2}{\Gamma\left(\frac{d-2}{2}-i q  \right)}\left( \frac{ r}{2\pi}  \right)^{\frac{d-2}{2}}
\int d\kappa\  \kappa^{-iq}e^{-i\kappa\cdot(\x-\eta)} K_{iq}(\kappa  r)
\label{FKsigwaves} \end{eqnarray} where $\kappa = \sqrt{\kappa\cdot\kappa}$. In
the second step we have used a well-known integral
representation of the Bessel-Macdonald function $K_{iq}(z)$
that we recall here for the reader's convenience:
\begin{equation}
K_{iq}(z)  = \frac12 \int_0^{\infty}  \frac{dR}{R} \  R^{- i q}
e^{- \frac12 z(R+\frac{1}{R} )} . \label{Bessrepr}
\end{equation} The function $K_{iq}(z)$ decreases exponentially
at large $z$; near the origin one has that \({K}_{iq}(z) \sim
\frac{\Gamma(iq)}{2}\left( \frac{z}{2}  \right)^{-iq}
+\frac{\Gamma(-iq)}{2}\left(\frac{z}{2}  \right)^{iq} \).
Therefore the integral (\ref{FKsigwaves}) converges at $\kappa = 0$
and provides a representation of ${\left({x \cdot \xi}   \right)
}^{- \frac{d-2}{2} + i q}$ for $\vert\Im q\vert <
{\frac{d-2}{2}}$.

\subsection{Expansion in spherical harmonics.}
There exists an abundant literature on Lobatchevski spaces that
is based on the use of generalized spherical harmonics in
connection with  spherical coordinate system. To render
possible a comparison of our results and methods with that
approach let us work out the change of basis.

By adopting the spherical coordinates of Section \ref{scsub} we
can write (see Eq. \ref{opq})
\begin{equation}e^{i R\,
{x\cdot\xi}}= e^{iR\cosh \psi}e^{-iR \sinh \psi \ {
\ndir_\phi}\cdot { \ndir_\theta}}.\end{equation} The spherical
leaves of our $(d-1)$-dimensional Lobatchevski space have
$(d-2)$ dimensions and we may expand the second factor at the
RHS in terms of generalized spherical harmonics $\Ylm({
\ndir_\phi})$  depending on $(d-2)$ angles, where $M$ is a
multi-index encoding  $(d-3)$ ``magnetic'' indices in addition
to $l$; in the standard two-dimensional case (that corresponds
to $d=4$) $\Ylm({ \ndir_\phi})$ are the standard spherical
harmonics $Y_{l,m}(\phi^1, \phi^2)$. The starting point is as
usual the well-known expansion of the exponential in terms of
Gegenbauer polynomials $\CG_l^\nu$ and Bessel functions (see
e.g. \cite{BAT} Eq. (7.10;5)):
\begin{equation}
 e^{i\gamma z} = \left(\frac2z  \right)^\nu \Gamma(\nu) \sum_{l=0}^{\infty}i^l (\nu+l)\CG_l^\nu(\gamma)J_{\nu+l}(z).
\end{equation}
The second step is to set $\nu$ to $\frac{d-3}{2}$ and take
advantage of the known (see e.g. \cite{Ratra85} Eq. (B.12))
expansion of $\CG_l^\nu({ \ndir_\phi}\cdot { \ndir_\theta})$ as
a sum of the generalized spherical harmonics:
\begin{equation}
e^{-iR \sinh \psi \ { \ndir_\phi}\cdot { \ndir_\theta}} = (2\pi)^\frac{d-1}{2}
\sum _{l=0}^\infty i^{-l} (R\sinh \psi)^{-\frac{d-3}{2}} J_{l+\frac{d-3}{2}} (R\sinh \psi)
\sum_{M} { \Ylm}({ \ndir}_\theta)  \Yslm({ \ndir}_\phi).
\end{equation}
By inserting this expression in the Euler representation
(\ref{comprep}) we get that
\begin{eqnarray}
  {\left({x \cdot \xi}   \right) }^{- \frac{d-2}{2} + i  {q}}\,& = &
    2 \pi \left(\frac{ 2\pi} {\sinh \psi}  \right)^\frac{d-3}{2} \;\;
  \sum _{l=0}^\infty \frac{{i^{-(l+\frac{d-2}{2} - i  {q})}}}{{\Gamma}\left(\frac{d-2}{2} - i  {q}  \right)} \;\;
 \sum_{M}  { \Ylm}({ \ndir}_\theta)  \Yslm({ \ndir}_\phi)
 \;\;
 \cr &&\int_0^\infty \frac{dR}{R}\
R^{\frac 12 - i  {q}}\, e^{i R\cosh \psi}
 J_{l+\frac{d-3}{2}}(R\sinh \psi)   .
  \label{comprep5}
\end{eqnarray}
The integral at the RHS is the Mellin transform of a product
that can be evaluated by the Mellin-Barnes integral. This is
 a way to directly check \cite{BAT}  Eq. (7.8;9):
\begin{eqnarray}
\int_0^\infty \frac{dR}{R}\ R^{\frac 12 - i  {q}}\, e^{i R\cosh
\psi} J_{l+\frac{d-3}{2}}(R\sinh \psi)= i^{l+\frac{d-2}{2}-iq}
\Gamma\left(l+\frac{d-2}{2}-i q  \right) P_{-\frac12+i q}^{-l-\frac{d-3}{2}}( \cosh \psi)
\end{eqnarray}
that holds for $\Re (l+\frac{d-2}{2}-i q ) >0$, that is for all
$l \ge 0$ provided\footnote{If we require $ {\left({x \cdot
\xi} \right) }^{- \frac{d-2}{2} - i {q}}$ to bear
simultaneously the same expansion, the condition of validity
becomes $\vert\Im q\vert < \frac{d-2}{2}$.}  $\Im q
> - \frac{d-2}{2}$. Therefore
\begin{eqnarray}
   \lefteqn{{\left({x(\psi,\phi) \cdot \xi(\theta)}   \right) }^{- \frac{d-2}{2} + i  {q}}\, =}\cr&&=
   2 \pi \left(\frac{ 2\pi} {\sinh \psi}  \right)^\frac{d-3}{2}
  \;\;  \sum _{l=0}^\infty \frac{ \Gamma\left(l+\frac{d-2}{2}-i q  \right) }  {\Gamma\left(\frac{d-2}{2} - i  q  \right)}
 P_{-\frac12+i q}^{-l-\frac{d-3}{2}}( \cosh \psi)
\;\; \sum_{M}  { \Yslm}({ \ndir}_\theta)  \Ylm({ \ndir}_\phi)= \cr&&=
a(q)
  \;\;  \sum _{l=0}^\infty
\;\; \sum_{M}  { \Yslm}({ \ndir}_\theta) Z_{iq,l,M}(\psi,{ \ndir}_\phi)
 \label{symsph}
\end{eqnarray}
where
\begin{equation}
a(q)= (2 \pi)^{\frac{d-1}2} \frac{\Gamma\left({-iq}  \right)}{\Gamma\left(\frac{d-2}{2}- iq  \right)} .
\label{adeq}
\end{equation}
and
\begin{equation}
Z_{iq,l,M}(\psi,{ \ndir}_\phi)=
   \frac{ \Gamma\left(l+\frac{d-2}{2}-i q  \right) }  {\Gamma\left(-i  q  \right)}\;\;
 {(\sinh \psi)^{-\frac{d-3}{2}}}\;\;P_{-\frac12+i q}^{-l-\frac{d-3}{2}}( \cosh \psi)
\;\;  \Ylm({ \ndir}_\phi).
\end{equation}
For $q\geq0$, the latter are the orthonormal eigenmodes (see
e.g. \cite{Ratra94}; note a slight modification in our
definition of $Z$) associated  with the spherical coordinates.

\subsection{Asymptotics.}
Using the coordinates (\ref{dscoor}) the "boundary" at infinity
of the manifold $\Lobad$ is attained for either $ r \to 0$ or
$ r \to \infty$ and the behaviour of the modes in these limits
is of importance. Actually, only the behaviour  at small $ r$
matters while the limit $ r\to \infty$ is rather an artifact of
the coordinate system (see Fig.   \ref{fig:xparam}). Eq.
(\ref{FKsigwaves}) together with the behaviour of
$K_{iq}(\kappa r)$ at small $ r$ give the following asymptotics:
\begin{equation}{\left({x( r,\x) \cdot \xi}   \right) }^{- \frac{d-2}{2} + i  {q}}\,  \sim
\  \,\,     r^{\frac{d-2}{2}- iq}
(\xi(\x)\cdot\xi)^{-\frac{d-2}{2}+ iq}\  \,\, + \  \,\, A(q)
\, r^{\frac{d-2}{2}+ iq} \delta(\xi(\x)-\xi). \label{psiorig}
\end{equation}
where
\begin{equation} A(q) =  \frac{(2\pi)^{\frac{d-2}{2}} 2^{-iq}
\Gamma\left({-iq}  \right)}{\Gamma\left(\frac{d-2}{2}- iq  \right)}. \label{Adeq}
\end{equation}

By sending the point $x$ that appears in Eq. (\ref{Sigwaves})
to the "boundary" at infinity of the manifold $\Lobad$ one gets
naturally the two-point kernel $(\xi\cdot
\xi')^{-\frac{d-2}{2}+ iq}$ on the asymptotic cone. As before,
this kernel admits  useful Euler and Fourier representations:
 \begin{eqnarray}
  {\left({\xi \cdot \xi'}   \right) }^{- \frac{d-2}{2} + i  {q}}\, &=&
   \frac{1}{{\Gamma}\left(\frac{d-2}{2} - i  {q}  \right)}\int _{0}^{\infty}\frac{dR}{R}\
 e^{-R\,  {\xi\cdot\xi'}}\,R^{\frac{d-2}{2} - i  {q}}\,
\label{Euconwaves}\\
& = &\frac{{2^{iq}\Gamma\left(
iq  \right)}}{{(2\pi)^{\frac{d-2}{2}}\Gamma}\left(\frac{d-2}{2} -
i{q}  \right)}
 \int d\kappa\  \kappa^{-2iq}e^{-i\kappa\cdot({\eta}-{\eta'})}
.\label{Fconwaves}\end{eqnarray} Let us discuss an immediate
application of this definition by establishing the relation
between modes with negative and positive values of $q$. Suppose
that $\left({x\cdot\xi}  \right)^{-\frac{d-2}{2} - iq}$ be
superposition of the modes
$\left({x\cdot\xi'}  \right)^{-\frac{d-2}{2} + iq}$, with $q>0$.

Homogeneity implies that
$\left({x\cdot\xi}  \right)^{-\frac{d-2}{2} - iq}$ and $\int
d\mu(\xi')\ \left({\xi\cdot\xi'}  \right)^{-\frac{d-2}{2} - iq}
\left({x\cdot\xi'}  \right)^{-\frac{d-2}{2} + iq}$ are
proportional. This can be explicitely shown by using the
integral representations (\ref{FSigwaves}) and
(\ref{Fconwaves}) in order to perform the latter integration:
\begin{equation}
\left({x\cdot\xi}  \right)^{-\frac{d-2}{2} - iq} = \frac{1}{A(q)} \int d\mu(\xi')\
\left({\xi\cdot\xi'}  \right)^{-\frac{d-2}{2} - iq}
\left({x\cdot\xi'}  \right)^{-\frac{d-2}{2} + iq}\label{qtominq}
\end{equation}

\section{Orthogonality of the modes on $\Lobad$ .\label{sec:scalprod}}

We are now ready to construct a basis for the Hilbert space
$L^{2}(\Lobad,d\mu)$ that can be used to study "ordinary"
quantum theories on the Lobatchevski space $\Lobad$. The word
ordinary refers to theories where the common wisdom and the
standard tools of quantum mechanics, including the
probabilistic interpretation, apply. As we have already said,
there are also non-standard theories, corresponding to the
allowed imaginary values of $q$. These theories will be
examined elsewhere.

\subsection{Scalar product of waves}
The modes (\ref{Sigwaves})  correspond to eigenvalues of the
continuous spectrum of the Laplacian. Of course modes
corresponding to distinct values of $q^2$ are orthogonal,
because of the self-adjointness of the Laplacian.  To find the
correct normalization  we study the distributional kernel
constructed by taking the scalar product of two modes
(\ref{Sigwaves}) for $q$ and $q'$ are real. It follows that
(see Section \ref{hhh})
\begin{eqnarray}
\FS_{q,q'}(\xi,\xi') &=& \int d\mu(x) \   {\left({x \cdot
\xi}   \right) }^{- {\frac{d-2}{2}} - i  {q}}\, \, {\left({x \cdot \xi'}   \right)
}^{- {\frac{d-2}{2}} + i  {q}'}= \cr  &=&  \frac{1}{\Nq(q)} \, \left[\delta(q-q') \,\delta(
\eta - \eta ') + \frac{1}{A(q)} \left(\xi\cdot\xi'  \right)^{-{\frac{d-2}{2}}-iq}\
\delta(q+q')  \right]
\label{scalpos2}
\end{eqnarray}
 while the normalization reads
\begin{equation}
 \frac{1}{\Nq(q)}= {a(q)a(-q)}=  {2\pi A(q) A(-q)} =
\frac
{(2\pi)^{d-1}\Gamma\left(iq  \right)\Gamma\left(-iq  \right)}
{\Gamma\left({\frac{d-2}{2}}+iq  \right)\Gamma\left({\frac{d-2}{2}}-iq  \right)}
.
\label{scalpos}
\end{equation}
We may recall  that $a(q)$ is given in (\ref{adeq}) and $A(q)$  in (\ref{Adeq}).
This result is indeed consistent with (\ref{qtominq}).

\subsection{Orthonormalized modes}

It is appropriate to introduce normalized modes and the
conjugate ones as follows. For $q \geq 0$, let us define
\begin{eqnarray} &&{\psi}_{iq} (x,\xi) =
\frac{\left({x \cdot \xi}   \right)^{- {\frac{d-2}{2}} +i
{q}}}{a(q)}  , \qquad {\psi}^*_{iq}(x,\xi) =
\frac{ \left({x \cdot \xi}   \right)^{- {\frac{d-2}{2}}
-i  {q}}}{a(-q)} \label{normmodes}
\end{eqnarray}
so that
\begin{equation}
\int d\mu(x) \   \psi^*_{iq}(x ,\xi ) \,{\psi}_{iq'}
(x,\xi')\, =
  \delta(q-q')\, \delta(\xi,\xi') .
\label{scalnorm} \end{equation} The set
$\{{\psi}_{iq}(x,\xi),\,\,q \geq 0,\;\xi \makebox{ on the
absolute}\}$ is then an orthonormal\footnote{From
(\ref{scalpos2}) it is however seen that the case $d=2$
exhibits some pecularities since the term proportional to
$\frac{1}{A(q)}$ does not vanish for $q=0$.} family of modes
for the Hilbert space $L^2(\Lobad,d\mu)$. The relation of these
waves with the more commonly  used waves in spherical
coordinates is given by Eq. (\ref{symsph}). The following
suggestive expansion in term of generalized spherical harmonics
is worth to be mentioned:
\begin{eqnarray}
\psi_{iq}
(x(\psi,\phi),\xi(\theta)) \, = \sum _{l=0}^\infty
 \sum_{M}  \Yslm(\ndir_\theta)  Z_{iq,l,M}(\psi,\ndir_\phi)
 \label{symsphexp}
\end{eqnarray}
and conversely
\begin{eqnarray}
Z_{iq,l,M}(\psi,\ndir_\phi)
 \, = \int  \Ylm(\ndir_\theta)\  \psi_{iq} \left(x(\psi,\phi),\xi(\theta)\right) d\mu(\theta)
 \label{symsphprod}
\end{eqnarray}
Our normalized plane waves are  therefore superpositions of the
spherical waves $Z_{iq,l,M}(\psi,\ndir_\phi)$ with weights
which are themselves normalized generalized spherical harmonics
evaluated at the direction of the vector $\xi(\theta)$
parametrizing the plane wave itself.

The advantage of the waves (\ref{normmodes}) is their
independence on the choice of particular coordinate systems
and, above all, their maximal symmetry. They really encode the
symmetry of the Lobatchevski space. Their representations in
terms of exponentials given in Sect. (\ref{sec:modereps}) also
renders feasible  calculations that are otherwise intractable
(see below and \cite{MSfollow}).

\subsection{Comments and details \label{hhh}}

\paragraph*{Factorization of $\FS_{q,q'}(\xi,\xi')$ .}

Let us insert in  Eq. (\ref{scalpos2}) the Fourier
representation (\ref{FSigwaves}) in the previous expression and
change to the variables $R = \frac{1}{u}$, $R' = \frac{1}{u'}$:
\begin{equation} \FS_{q,q'}(\xi,\xi') = \frac{\int \frac{du}{u} \  \frac{d
u'}{u'} \,\frac{d r}{ r} \,d\kappa \   r^{i(q-q')}{u}^{-
i{q}}\,{u'}^{ iq'}  \, e^{i\kappa \cdot (\eta-\eta')} \, e^{-\frac{u
+u'}{2}\kappa^2 -\frac{(u+u')}{2uu'} r^2}}{{\Gamma}\left({\frac{d-2}{2}}
+iq  \right){\Gamma}\left({\frac{d-2}{2}} - iq'  \right)}.
\end{equation}
The integral over $\kappa$ can be factorized by the changes $ r^2 =
\sigma^2 uu' \kappa^2$, $u= \frac{v}{\kappa^2}$ and $u'=
\frac{v'}{\kappa^2}$:
\begin{eqnarray}
&& \FS_{q,q'}(\xi,\xi')  = \frac{\Delta_{q,q'}}{{\Gamma}\left({\frac{d-2}{2}}
+iq  \right){\Gamma}\left({\frac{d-2}{2}} - iq'  \right)}   \   \int d\kappa \
\kappa^{i{(q-q')}} e^{i\kappa \cdot (\eta-\eta')} \, \label{aa}\\&& \cr &&
\Delta_{q,q'} = \int \frac{d\sigma}{\sigma} \ \sigma^{i{(q-q')}}\int
\frac{dv}{v}\  {v}^{-i\frac{q+q'}{2}}\, e^{-\frac{v}{2}
-\frac{v}{2}\sigma^2}\, \int \frac{dv'}{v'}\  {v'}^{
i\frac{q+q'}{2}}\, e^{-\frac{v'}{2} -\frac{v'}{2}\sigma^2} .
 \label{scalrep}
\end{eqnarray}

\paragraph*{Evaluation of $\Delta_{q,q'}.$}
\paragraph*{1) $q+q'\not = 0$.} The second and third integrals at RHS
can be evaluated as follows:
\begin{equation}
\int \frac{dv}{v} \  {v}^{- i\frac{q+q'}{2}}\, e^{-\frac{v}{2}
-\frac{v}{2}\sigma^2}\, = {2^{-\frac{iq+iq'}{2}}}{{\left( 1 +
\sigma^2   \right) }^\frac{iq+iq'}{2}}
\Gamma\left(-\frac{iq+iq'}{2}  \right).
\end{equation}
The integral converges only if $\Im(q+q') >0$. Otherwise it is
defined in a generalized sense as a meromorphic function of the
complex $(q+q')$ variable. It then follows that
\begin{equation}
\Delta_{q,q'}  =  {\Gamma}\left(-iq   \right){\Gamma}\left({i }q
  \right)\,2\pi\delta(q-q') .
\end{equation}
\paragraph*{2) $q-q' \not = 0$.} In this case we can extract a finite
contribution by exchanging the integration order in Eq.
(\ref{scalrep}):
\begin{eqnarray} && \int
\frac{d\sigma}{\sigma} \  \sigma^{i{(q-q')}}\;
e^{-\frac{v+v'}{2}\sigma^2}=\frac12\left(\frac{v+v'}{2}  \right)^{-\frac{iq- iq'}{2}}\,
 \Gamma\left(\frac{iq- iq'}{2}  \right) .
 \end{eqnarray}
 This expression is valid  for $\Im (q-q') < 0$. Otherwise it
is defined in a generalized sense as a meromorphic function of
the complex $(q-q')$ variable. By introducing the variables  $
\lambda = \frac{v'}{v}$ and $\mu = \frac{v+v'}{2} $ Eq.
(\ref{scalrep}) becomes
\begin{equation}
 \Delta_{q,q'}  = \frac1{2}\;\Gamma\left(\frac{iq-iq'}{2}  \right)
\int \frac{d\lambda}{\lambda} \  \frac{d \mu}{\mu}   \
{\mu}^{-\frac{iq-iq'}{2}}{\lambda}^{ \frac{iq+iq'}{2}} e^{-\mu}
 = {\Gamma\left(iq  \right)  \Gamma\left(-iq  \right)}
 \,\,2\pi\, \delta(q+q').
\end{equation}
Gathering all terms together finally yields
\begin{equation}
 \Delta_{q,q'}  =
 2\pi{\Gamma\left(iq  \right)  \Gamma\left(-iq  \right)}
 \, \left( \delta(q-q') + \delta(q+q')  \right).
 \label{Deltaosos}
\end{equation}
and Eq. (\ref{scalpos2}) follows from Eq. (\ref{aa}).

\section{Sketch of an integral transformation theory. Completeness. \label{sec:inttrans}}

\subsection{Transform and inversion. \label{sect:scaltrans}}
Let $f(x)$ be a smooth function defined on $\Lobad$. We define
the following transform \cite{gelfand}:
\begin{equation}
f(x)\to \tilde{f}(\xi,q)  = \int d\mu(x)  \  \left({x \cdot \xi}
  \right)^{- \frac{d-2}{2} - i {q}}\, \, f(x) .
 \label{scalcoeff}
\end{equation}
Variables paired by the transform are
\begin{equation} \Lobad \ni x
\longleftrightarrow (\xi,q) \in C^+\times {\mathbb R}.
\label{xtoxiq}
\end{equation}
Since  the transformed function is an homogeneous function of
the $\xi$ variable, it is enough to take $\xi$ on the absolute
and therefore there are $(d-1)$ dimensions on both sides of
this pairing. The aim is now to reconstruct the function $f(x)$
in terms of $\tilde{f}(\xi,q) $ by inverting the previous
transform. The inversion formula we will heuristically prove in
this section is the following\footnote{Another choice would be
to define a normalized transform:
\begin{equation}
f(x) \to \tilde{f}(\xi,q)  = \int d\mu(x) \ {\psi}^*_{iq}
(x,\xi)\, f(x) . \nonumber
\end{equation}
In this case the inversion formula would have unit weight
\begin{equation} f(x) = \int_{0}^{\infty} dq \  \int d\mu(\xi)
\   \psi_{iq}(x,\xi) \tilde{f}(\xi,q). \nonumber
\end{equation}
Our above definition (\ref{scalcoeff}) follows similar choices
of normalization such as the Mehler-Fock or the
Kontorovich-Lebedev transforms, see e.g. \cite{BAT},
Eqs. (3.15; 8-9). \label{foot:normscaltrans} }:

\begin{eqnarray} f(x) &=& \int_{0}^{\infty}  {\Nq(q)}{dq }\   \int d\mu(\xi)
 \  \left({x \cdot \xi}   \right)^{- \frac{d-2}{2} +i  {q}}
\tilde{f}(\xi,q)  \label{inv}\end{eqnarray} where $\Nq(q)$ is given
in Eq. (\ref{scalpos})

\subsection{Completeness. \label{sec:compl}}

Let us consider the (formal) integral operator
\begin{equation} \DEL(x,x') = \int_{0}^{\infty} dq' \ \int
d\mu(\xi)  \  {\psi}_{iq'} (x,\xi) {\psi}^*_{iq'} (x',\xi)
\label{DelL2} \end{equation}    in the Hilbert space
$L^2(\Lobad,d\mu)$. Using the coordinates (\ref{dscoor}), this
Hilbert space may be concretely realized as a tensor product:
\begin{equation}
L^2(\Lobad,d\mu) =L^2({\mathbb R}^+\times {\mathbb
R}^{d-2},{ r^{1-d}} {d r} \, d\x)= L^2({\mathbb R}^+,{ r^{1-d}}
{d r})\otimes L^2({\mathbb R}^{d-2},d\x). \label{hilbertspace}
\end{equation}
The space $L^2(\Lobad,d\mu)$ is therefore generated by  finite
linear combinations of factorized functions $f( r,\x) =
g( r)h(\x)$ where the factors are such that $g( r)\in
L^2({\mathbb R}^+,{ r^{1-d}} {d r})$ and $h\in L^2({\mathbb
R}^{d-2},d\x)$. Consider one function of this type. Using the
integral representation (\ref{FSigwaves}) we can express the
operator (\ref{DelL2}) as follows:
\begin{eqnarray} \int  d\mu(x') \  \DEL(x,x')f(x')   = \frac{ r^{\frac{d-2}{2}} }{(2\pi)^{d-2}}\int d \kappa\  \int
d\x' \  e^{i \kappa\cdot\left({\x}-{\x'}  \right)}h(\x')
\,\cr \ \ \ \ \ \ \ \ \ \ \ \ \ \ \times \int_{0}^{\infty}dq \  \frac{2 }{\pi\Gamma\left(i q  \right)\Gamma\left(-i q  \right)} K_{iq}(\kappa
 r)\int_0^\infty \frac{d r'}{ r'} \   K_{iq}(\kappa
 r')\frac{g( r')}{{ r'}^{\frac{d-2}{2}}}.  \label{fkl} \end{eqnarray}

The function ${{ r}^{-\frac{d-2}{2}}}\,{g( r)} \in L^2({\mathbb
R}^+,{ r^{-1}} {d r})$ and this assures convergence of the
inner integral. The integral over $ r'$ and $q$ are then just
an instance of the Kontorovich-Lebedev \cite{davis} inversion
formula, that holds true for quite general classes of functions
and distributions:
\begin{eqnarray} \frac{g( r)}{{ r}^{\frac{d-2}{2}}}&=& \int_{0}^{\infty}dq  \ \frac{2 }{\pi\Gamma\left(i q  \right)\Gamma\left(-i q  \right)} K_{iq}(\kappa
 r)\int_0^\infty \frac{d r'}{ r'} \   K_{iq}(\kappa
 r')\frac{g( r')}{{ r'}^{\frac{d-2}{2}}}.
\end{eqnarray}
The remaining integral in Eq. (\ref{fkl}) is then just Fourier
inversion formula. Taking finite linear combinations  we
finally get
\begin{eqnarray} &&\int  d\mu(x') \  \DEL(x,x')f(x')  = f(x)
\label{delsupp}
\end{eqnarray}
on (a dense subset) of $L^2(\Lobad, d\mu)$. This shows the
validity of the inversion formula (\ref{inv}).

\section{Projectors. Representations of the principal series. \label{sec:projq}}

At this point we may  introduce the integral kernels
\begin{equation}\PR_q(x,x') = \int d\mu(\xi)\  {\psi}_{iq} (x,\xi)
{\psi}^*_{iq} (x',\xi). \label{projq} \end{equation} It is
immediately seen that the kernels $\PR_q$ satisfy the following
projector relations:
\begin{equation}
\int d\mu(x'') \   \PR_q(x,x'')\PR_{q'}(x'',x')= \delta(q-q')\
\PR_q(x,x'). \label{repker}
\end{equation}
The operator $\PR_q(x,x')$ is  the projector on the subspace of
a given $q^2$ and as such $\PR_q(x,x')$ is a positive-definite
kernel. Starting from the projector  $\PR_q(x,x')$ we can
construct a representation of the invariance group of $\Lobad$
in the usual way: let us consider the space of smooth rapidly
decreasing functions on ${\cal S}(\Lobad)$ endowed with the
left regular action $(T_g f)(x) = f(g^{-1}x)$, $g\in
SO_0(1,d-1)$. As usual, let us introduce in such a space the
scalar product
\begin{equation}
\langle f, f' \rangle = \int_{\Lobad} f{^*}(x)\PR_q(x,x') f'(x')
dx dx'
\end{equation}
${\cal S}(\Lobad)$ is a pre-Hilbert space. By quotienting and
completing we obtain a Hilbert space carrying an irreducible
unitary representation of the Lorentz group labeled by the
real, non-negative parameter $q$. The set of such
representations is called the principal series.

Consider now any invariant (possibly positive-definite)
two-point kernel $W(x,x')$ on the hyperboloid $\Lobad$. If we
assume suitable growth properties of  $W$ at infinity we may
expect that it can be decomposed as a superposition of the
projectors $\PR_q(x,x')$ (representations of the principal
series):
\begin{equation}
W(x,x') = \int_0^\infty  \rho(q) \PR_q(x,x') dq.
\end{equation}
In particular the kernel $ \DEL(x,x')$, as defined in Section
(\ref{sec:compl}), determines the standard $L^2$ Hilbert
product on the hyperboloid $\Lobad$ and the so-called regular
representation:
\begin{equation}
\DEL(x,x')  = \int_0^\infty  \PR_q(x,x') dq
\end{equation}
which can be equivalently viewed as the decomposition of the
regular representation into representations of the principal
series  (Plancherel's formula):
\begin{eqnarray}\int d\mu(x) \  f{^*}(x) \, g(x)
= \int_{0}^{\infty}   {\Nq(q)} {dq}    \ d\mu(\xi)
\tilde{f}^*(\xi,q) \tilde{g}(\xi,q) . \label{planch}
\end{eqnarray}

\paragraph*{Evaluation of  $\PR_q(x,x')$.} The explicit evaluation of
the integral (\ref{projq}) is most easily done by integrating
on the spherical basis (\ref{sphericcone}) of the absolute
($\xi^0=1$). The result must depend only on the scalar product
$x\cdot x'$: without loss of generality we may choose $x =
(1,0,\ldots,0)$ and $x'=(\cosh\phi,-\sinh\phi,\ldots,0)$ so
that $x\cdot x' = \cosh\phi. $. Since  $x \cdot \xi = 1,$ and
$x' \cdot \xi = \cosh\phi+\cos \theta_1 \sinh \phi$, the
integral (\ref{projq}) becomes
\begin{eqnarray}
\frac{2\pi^{\frac{d-1}{2}}}{\Gamma\left(\frac{d-1}{2}  \right)} \Nq(q)
\int_0^\pi d\theta_1 \  \left(\cosh\phi+\cos \theta_1 \sinh
\phi  \right)^{- \frac{d-2}{2} -iq}(\sin \theta_1)^{d-3} = \cr=
 \frac{2\pi^{\frac{d-1}{2}}} {\Gamma\left(\frac{d-1}{2}  \right)}
\Nq(q) \Gamma\left(\frac{d-1}{2}  \right)
2^{\frac{d-3}{2}}(\sinh\phi)^{-\frac{d-3}{2}}P^{-\frac{d-3}{2}}_{-\frac{1}{2}+iq}(\cosh\phi)
\end{eqnarray}
so that the final result reads
\begin{equation} \PR_q(x,x') =  \omega_{d-1}\, \Nq(q)  \,
{P}^{(d)}_{-\frac{d-2}{2} + i{q}}(x \cdot x'). \label{projqexpl}
\end{equation}
The factor $\omega_{d-1}$ is the hypersurface of the sphere
${\Bbb S}^{d-2}$ (see Footnote   \ref{foot:d-spheres}) and $
\Nq(q)$ is given by (\ref{scalpos}).  The result is expressed
in terms of the so-called generalized Legendre function:
\begin{equation}
{P}^{(d)}_{-\frac{d-2}{2} + i{q}}(z )  =   2^{\frac{d-3}{2}}\,
\Gamma\left(\frac{d-1}{2}  \right)\,(z ^2-1)^{-\frac{d-3}{4}}\,
P^{-\frac{d-3}{2}}_{-\frac{1}{2} + i{q}}(z ) \label{Pdef}
\end{equation}
where $P^\mu_\nu(z)$ denotes the usual Legendre function of the
first kind, defined and one-valued in the complex $z$-plane cut
on the reals from $-\infty$ to 1 \cite{BAT}. The  function $(z
^2 -1)^\alpha$ is defined and one valued on the same cut
complex plane (with the natural definition for real $z  >1$) so
that the function ${P}^{(d)}_{-\frac{d-2}{2} + i{q}}(z )$ is
regular at $z = 1$ and its cut goes from $z=-\infty$ to $z=-1$
only.

\section{Milne's universe. \label{sec:milne}}
 As a first application of the general construction displayed in the previous sections,
 we  discuss here Quantum Field Theory in the universe of Milne.

\begin{figure}[h]
\begin{center}
\includegraphics[height=9cm]{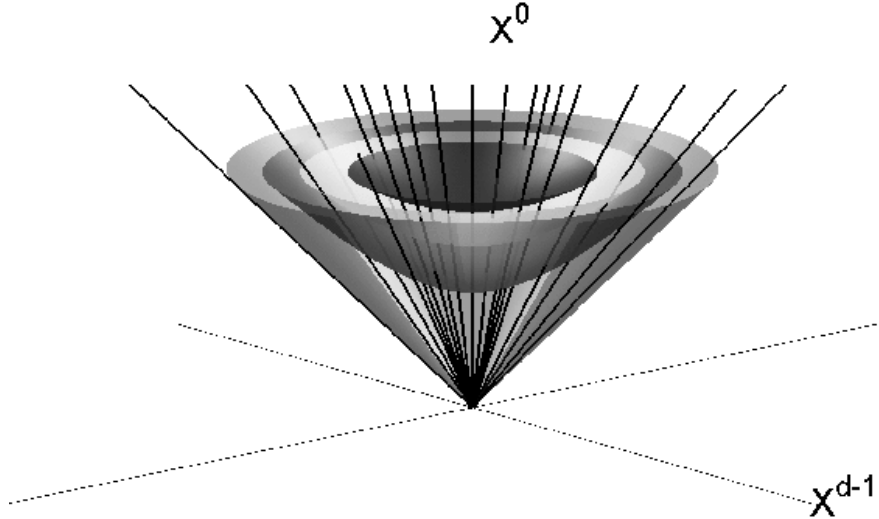}
\caption{\em A view of the Milne universe.  The spacetime
curvature of this model is zero. Surfaces of constant time are
copies of the Lobatchevski space $\Lobad$. The straight lines represented in figure are
the geodetic wordlines of particles having constant spatial coordinates.} \label{milnefig}
\end{center}
\end{figure}

Milne's universe is a simple model of an expanding universe
obtained as a solution of the Einstein equations in vacuo with
zero spacetime curvature and nonzero spatial curvature. There
is however no need of General Relativity to talk about this
model: Fig.   \ref{milnefig} shows how the Milne universe can be
constructed as a foliation with Lobatchevskian leaves of the
interior of the the future cone of an event (the "Big Bang") of
a Minkowski spacetime. A quantitative description is very easy:
let $X^\mu$ denote the coordinates of an event $X$ of a
($d$-dimensional) Minkowski spacetime. Consider the future cone
of  the origin of the chosen inertial system (as in Fig.
  \ref{milnefig})  and introduce there the  noninertial
coordinate system:
\begin{equation}
X^{\mu}({t},x) = t \, x^\mu ,\;\;\mu = 0,\ldots,d-1,
\label{dScoorgen}
\end{equation}
where $x\cdot x = 1$, i.e. $x\in \Lobad$. Milne's line element
is simply  Lorentz invariant interval of the ambient spacetime
expressed in the coordinates (\ref{dScoorgen}):
\begin{eqnarray}ds^2 &=&    {\left(dX^{0}  \right)}^2 -
{\left(dX^{1}  \right)}^2-\ldots - {\left(dX^{{d-1}}  \right)}^2 =
d{t}^2 - t^2 \, dl^2_{d-1}.
\end{eqnarray}
Milne's universe has therefore the structure of a warped
product of a half line (the cosmic time) times the Riemannian
manifold $\Lobad$; the warping function is just the cosmic time
$t$ (see e.g \cite{Bertola}).

As old as it is, this model has never become obsolete and
disappeared from the scientific and cosmological debate; its
predictions are in surprisingly good agreement with the current
cosmological observations \cite{Kutschera}. A recent appearance
of  Milne's model in M-theory is also worth to be mentioned
\cite{Khoury}.

The simple question we ask ourselves here is that of finding
the expansion of the exponential plane waves of the Minkowski
spacetime on the base of modes (\ref{normmodes}). This is a
preliminary step to study quantum theories on the Milne
universe in much the  same way as finding an expansion of plane
waves in spherical harmonics is a starting point in studying
spherical symmetric potentials in ordinary quantum mechanics.
As an immediate bonus of our approach there is an easy
construction the Wightman vacuum in Milne's coordinates. We
give here a new and we believe simple approach to solve this
old problem \cite{disessa}. Our approach may also be used to
study other vacua, as for instance the thermal vacuum at a
given temperature.

\subsection{Free fields}
To fix ideas and notations, and also to put the results in
perspective, it is useful to begin by shortly reviewing the
theory of a Klein-Gordon quantum field of mass $m$ on a
$d$-dimensional Minkowski spacetime ${\mathbb M}^{d}$ with
inertial coordinates $(X^0,X^1,$ $\ldots,X^{d-1})$:
\begin{equation}
 \left(\square + m^2   \right)\phi=
0. \label{kkgg}
\end{equation}
It is enough to solve the Klein-Gordon equation for the
two-point vacuum expectation value
\begin{equation}
W_m(X-Y) = \langle \Omega, \phi(X) \phi(Y) \Omega \rangle.
\end{equation}
The truncated $n$-point functions are assumed to vanish and the
two-point function encodes all the information necessary to
fully reconstruct the theory. Actually, for Klein-Gordon fields
satisfying the Wightman axioms the vanishing of the truncated
$n$-point functions is not an assumptions but it is a result
\cite{Streater} and a Klein-Gordon field is necessarily free.
Eq. (\ref{kkgg}) is most easily solved in Fourier space, where
it becomes algebraic:
\begin{equation}
(p^2-m^2)\tilde W_m(p) = 0 .
\end{equation}
There are infinitely many inequivalent solutions and a
criterium is to be found to select one among them. Assumption
of positivity of the spectrum of the energy-momentum operator
is the most popular possibility and leads to
\begin{equation}
\widetilde W_m(p) \simeq \theta(p^0) \delta(p^2-m^2), \label{wight}
\end{equation}
where $\theta(p^0)$ denotes Heaviside's step function.
Inversion gives
\begin{equation}
W_m(X,X')= \langle \Omega, \phi(X)\phi(X')\, \Omega \rangle =
\frac{1}{(2\pi)^{d-1}}\int _{{\mathbb R}^{d}}  dp \ e^{-i p\cdot (X-X')}
\theta(p^0)\delta(p^2-m^2)  . \label{01}
\end{equation}
The choice made in Eq. (\ref{wight}) selects the Wightman
vacuum $\Omega$ which is uniquely characterized by the
positivity of the spectrum of the energy operator in any
Lorentz frame. This property is equivalent to certain
analyticity properties of the correlation functions that can be
deduced by direct inspection of Eq. (\ref{01}). One sees that
the Wightman function  $W$ can be uniquely extended to a
function holomorphic in the past tube $T^{-}$ as a function of
the difference variable  $(X-X')$ where
\begin{equation}T^-= \{ X+iY ,\;\; Y^2>0, Y_0<0\}. \label{MinkTmoins}\end{equation}
If we consider the plane waves on the mass shell we see that
positive frequency waves $\exp(-i\sqrt{p^2+m^2} X^0 + i
\vec{p}\cdot{\vec X})$ admit a natural continuation in the past
tube where they are decreasing  while negative frequency waves
$\exp(i\sqrt{p^2+m^2} X^0 - i \vec{p}\cdot{\vec X})$ may be
considered for complex events belonging to the future tube
\begin{equation}T^+ = \{ X+iY ,\;\; Y^2>0, Y^0>0\}.\label{MinkTplus} \end{equation}
These properties are the link between the standard choices in
the canonical quantization procedure and the analyticity
structure of the waves and the Wightman function.

\subsection{Plane waves: projection and inversion. QFT. \label{sec:Milinv}}

Let us consider therefore a Minkowskian plane wave  on the mass
shell $p^2 = m^2$, $p^0>0$  written in Milne's coordinates:
\begin{equation} \exp({ip\cdot X}) = \exp{(i t\,
p\cdot x)}. \label{Milwave}\end{equation}  The wave is
naturally extended in the future tube $T^+$;  in particular we
will consider the complex events
\begin{equation}
Z^{\mu}(\tau,x) = \tau\, x^\mu ,\;\;\;\tau = t+is
\label{Miltube},\;\;\;\Im \tau =  s>0,
\end{equation} that  belong to the future tube. Similarly the wave $\exp \left(-i \tau p\cdot
x  \right)$ is naturally extended to the past tube and in
particular we will consider the events $Z^{\mu}(\tau,x)\in T^-$
 that belong in the past tube  for $\Im \tau<0$.

We now look for an expansion of the Minkowskian plane waves
adapted to Milne's geometry. As explained in the previous
sections, the first step is to compute their scalar products
with the modes (\ref{Sigwaves}) as follows:
\begin{equation}
F^{\pm}_{q}(\tau,\xi,p)
\int_{\Lobad} d\mu(x) {\ }
\left({x\cdot\xi}  \right)^{-{\frac{d-2}{2}} - iq} e^{\pm i\tau\,p\cdot x} .
\label{Milscal}
\end{equation}
Here $F^{\pm}_q$ are defined respectively for $X(\tau,x) \in
T^\pm$, i.e. $\Im \tau > 0$ resp. $\Im \tau < 0$. The Lorentz
invariance of the measure implies that $F^{\pm}_q$ may depend
only on the invariant $(\xi\cdot p)$. Homogeneity of the
integrand then gives that
 \[F^{\pm}_{q}(\tau,\xi,p) = f^{\pm}_q(\tau)\left({\xi\cdot
p}  \right)^{-{\frac{d-2}{2}} - iq}.\] The steps to explicitly compute the
function $f$ are summarized at the end of this section; here is
the result:
 \begin{equation}
 F^+_{q}(\tau',\xi,p) =   i\pi \left(\frac{2 \pi i}{m\tau'}
  \right)^{\frac{d-2}2} \left(\frac{p \cdot\xi}m  \right)^{-{\frac{d-2}{2}} - iq}
e^{-\pi q}H^{(1)}_{i q}\left(m\tau'  \right),\;\;\;\;\Im\tau'>0
\label{Milscalres}
\end{equation}
\begin{equation}
F^-_{q}(\tau,\xi,p) =  \frac{\pi}{i}
\left(\frac{2 \pi}{im\tau}   \right)^{\frac{d-2}2} \left(\frac{p
\cdot\xi}m  \right)^{-{\frac{d-2}{2}} - iq} e^{\pi q} H^{(2)}_{i
q}\left(m\tau  \right),\;\;\;\;\;\;\;\Im\tau<0 \label{Milscalstarres}
 \end{equation}
Note that $H^{(2)}_{i q}\left(m\tau  \right) \propto e^{-im\tau}$
when $\tau \to \infty$. As expected \cite{disessa} the Hankel
function $H^{(2)}_{i q}$ plays the role of the positive
frequency solution of the Klein-Gordon equation when separated
in the coordinates (\ref{dScoorgen}). The result in our
construction  comes out automatically from the known
analyticity properties (\ref{MinkTmoins}) of the Minkowskian
waves.

Inversion is obtained by means of Eq. (\ref{inv}); this yields
the expansion of the exponential plane wave (\ref{Milwave}) in
terms of the wavefunctions (\ref{Sigwaves}); the $(d-1)$
parameters $p$ are  described by the $(d-2)$ degrees of freedom
of $\xi$ on the absolute plus one degree of freedom of the $q$
variable:
\begin{eqnarray}
\exp{i \tau(p\cdot x)} &=&
i\pi \left(\frac{2 \pi i}{m\tau}  \right)^{\frac{d-2}2}
 \int_{0}^{\infty}  {\Nq(q)}{dq }\ e^{-\pi q}H^{(1)}_{i q}\left(m\tau  \right)  \cr && \times\int d\mu(\xi) \
  \left(\frac{p \cdot\xi}m  \right)^{-{\frac{d-2}{2}} - iq}
 \left({x \cdot \xi}   \right)^{- \frac{d-2}{2} +i  {q}}
 \label{milwaveexp}
\end{eqnarray}
and similarly for the other wave. The integration over the
absolute at the RHS can be performed (see Eqs. (\ref{projq})
and (\ref{projqexpl})) and there results a one-dimensional
integral expansion over the projectors $\PR_q$ as follows:
\begin{eqnarray}
 e^{i\tau(p\cdot x)}&=&
i\pi \left(\frac{2 \pi i}{m\tau}   \right)^{\frac{d-2}2} \int_0^\infty {dq} \
e^{-\pi q}H^{(1)}_{i q}\left(m\tau  \right) \PR_q\left(\frac{p\cdot
x}{m}  \right),\,\,\Im \tau>0,
\cr
e^{-i\tau(p\cdot x)}&=& -i\pi  \left(\frac{2 \pi }{im\tau}   \right)^{\frac{d-2}2}
\int_0^\infty
{dq} \  e^{\pi q}H^{(2)}_{i q}\left(m\tau  \right)
\PR_q\left(\frac{p\cdot x}{m}  \right),\,\,\Im \tau<0
.
\label{milinv}
\end{eqnarray}
The details of the calculation are given at the end of the
present section. As an immediate bonus these formulae provide
an expansion of the Wightman canonical Klein-Gordon quantum
field theory expressed in Milne's coordinates. Indeed, since
the theory is completely encoded in the two-point function and
the invariant measure on the mass shell is proportional to the
invariant measure $d\mu$ on $\Lobad$, we can insert Eqs.
(\ref{milinv}) into Eq. (\ref{01}) and use Eq. (\ref{repker})
to get
\begin{eqnarray}
W^{}_m(X,X') &=&  \frac{1}{(2\pi)^{d-1}}\int _{{\mathbb R}^{d}} dp \  e^{-i
p\cdot (X-X')} \theta(p_0)\delta(p^2-m^2) \ \cr &=&
\frac14 {\pi}
{(\tau \tau')} ^{-\frac{d-2}2}\int \ {dq} \   H^{(2)}_{i
q}\left(m\tau  \right) H^{(1)}_{i q}\left(m\tau'  \right)
\PR_{q}\left({x\cdot x'}  \right), \label{MilneWexp}
\end{eqnarray}
where $\Im\tau <0, \Im \tau'>0$. As a verification, let us show
that the theory is indeed canonical by  computing the following
equal time commutation relations:
\begin{eqnarray}
&& [\phi(t,x),\pi(t,x')] = \cr
&& = \frac {\pi m t}4 \int
{dq} \   \left(H^{(2)}_{i q}\left(mt  \right) {H^{(1)}_{i q}}'(mt) -
{{H^{(2)}_{i q}}'\left(mt  \right)} {{H^{(1)}_{i
q}}}(mt)  \right) \PR_{q}\left({x\cdot x'}  \right)\cr
 &&  =   i  \ \delta(x,x'),
\label{Milcomm}
\end{eqnarray}
where $ \delta(x,x')$ is understood in the sense of Section
(\ref{sec:compl}).

\subsection*{Comments and details}

\paragraph*{Evaluation of $F_q$.}

In this section we show that the expressions of $F_q$ as given
in Eqs. (\ref{Milscalres}) and (\ref{Milscalstarres}) hold
true. To this purpose, let us parametrize $x$ as in Eq.
(\ref{dscoor}) and similarly write the momentum vector $p$ as
follows:
\begin{equation}
p = \left\{\begin{array}{lclcl}
{p^{0}} &=& \frac m{2\lambda}  (1+\kappa^2 +\lambda^2) \\
{p^{i}}  &=&  \frac m{2\lambda} {\kappa^i}\\
{p^{d-1}}  &=&\frac m {2\lambda} ({1-\kappa^2 -\lambda^2})
\end{array}   \right.
\label{ka}.
\end{equation}
This yields
\begin{equation}
\exp i p\cdot X=\exp i\tau p\cdot x =\exp \frac
{im\tau}{2 r\lambda}\left[{(\x - \kappa)^2 +  r^2 + \lambda^2}   \right].
\label{explicit2}
\end{equation}
By using the integral representation (\ref{comprep}) and
performing the Gaussian integral one gets
\begin{equation}
F^+_{q}=\frac{(2 \pi  )^{\frac{d-2}2}{i^{-\frac{d-2}{2} -i \,q}}}
{{\Gamma}\left(\frac{d-2}{2} + i \,q  \right)}
\int \frac{dR}{R}{\ } \frac{{R}^{\frac{d-2}{2} + i q}}
{[-i(T+R)]^{\frac{d-2}2}}\int \frac {dr}{r} {\ }  r^{-\frac{d-2}2}
 e^{ \frac {ir (R+T)}{2}
+\frac {i {\lambda^2}T}{2r}+i\frac{T R (\kappa-\eta )^2}{2 r
(T+R)}}
\label{Milscal3}
\end{equation}
where the integral over $R$ is along an arbitrary straight
half-line such that $0<\Arg (R)<\pi$; to simplify notations we
have put
\begin{equation}
T = \frac m \lambda(t+is)=\frac {m \tau}\lambda.\label{TT}
\end{equation} The
evaluation of the remaining integrals is simplified by the
introduction of a new complex variable replacing the
$r$-coordinate: given $R$ and $T$ in the upper complex plane we
define
\begin{equation}
v  = \left({\frac{1}{R} + \frac{1}T}   \right)r,
\;\;\;\;-\pi<\Arg(v)<0. \label{path}\end{equation}
Let us use
the freedom in choosing the integration path in the complex
$R$-plane and take $\Arg(R) = \Arg(T)$ (path $\gamma$); it
follows that $\Arg(v) = -\Arg(R)$ (path $\hat\gamma$) and the
previous expression becomes
\begin{eqnarray}
F^+_{q}=\frac{i^{-\frac{d-2}{2} -i \,q}}
{{\Gamma}\left(\frac{d-2}{2} + i q  \right)} \left(\frac{2i\pi}{T}   \right)^{\frac{d-2}2}
\int_\gamma \frac{dR}{R}{\ }  {R}^{i
q}\int_{\hat\gamma} \frac {dv}{v} {\ } {v}^{-\frac{d-2}2}
 e^{ \frac {iRTv}{2}
+\frac {i {\lambda^2}(R+T)}{2{R}v}} {e^{\frac{i(\kappa -\eta)^2}{2 v}}
}.
\end{eqnarray}
In this expression we interchange the integration order,
introduce the real variable $S = Rv$ and the inversion $u =
\frac{\lambda}{v}$ (that implies $0<\Arg(u)=Arg(R)<\pi$):
\begin{equation}
F^+_{q}=\frac{(i\lambda)^{-\frac{d-2}{2} -i \,q}}{{\Gamma}\left(\frac{d-2}{2} + i q  \right)}
\left(\frac{2i\pi}{T}   \right)^{\frac{d-2}2}
 \int_{\gamma}
\frac {du}{u}{\ }  {u}^{\frac{d-2}2+iq}{e^{\frac{i(\kappa -\eta)^2
+i\lambda^2}{2 \lambda}u} } \int_0^\infty \frac{dS}{S}{\ }  {S}^{i q} \,
 e^{ \frac {iST}{2}+\frac {i {\lambda^2}T}{2S}}. \label{path2}
\end{equation}
In the previous expression we recognize (see Eq.
(\ref{comprep})) the scalar
\begin{equation}
 \left(\frac{p \cdot\xi}m  \right)^{-{\frac{d-2}{2}} - iq} = \frac{ i^{-{\frac{d-2}{2}}-iq}}{{\Gamma}\left({\frac{d-2}{2}} + i\,q  \right)}
 \int_{\gamma}\frac{du}{u} {\ } u^{{\frac{d-2}{2}} + i \,q} e^{iu\,\left[\frac{\left({\kappa}-{\eta}  \right)^2}{2\lambda} + \frac{\lambda}{2}  \right] }
\end{equation}
and the representation (\ref{Bessrepr}) of the Bessel function
$K_{i q}(z)$.  Taking into account Eq. (\ref{TT}) and the phase
of $\tau$ we get
\begin{eqnarray}
F^+_{q}(\tau,\xi,p)&=&  2 \left(\frac{2 \pi i}{m\tau}
  \right)^{\frac{d-2}2} \left(\frac{p \cdot\xi}m  \right)^{-{\frac{d-2}{2}} - iq}
K_{i q}\left(-im\tau  \right) \cr
&=&  i\pi \left(\frac{2 \pi i}{m\tau}
  \right)^{\frac{d-2}2} \left(\frac{p \cdot\xi}m  \right)^{-{\frac{d-2}{2}} - iq}
e^{-\pi q}H^{(1)}_{i q}\left(m\tau  \right),\;\;\;\;\Im\tau>0.
\end{eqnarray}
This is the result given in Eq. (\ref{Milscalres}). The
expression (\ref{Milscalstarres}) is obtained by complex
conjugation, as can be readily inferred from (\ref{Milscal}).
This completes the proof of the expansions given in Eqs.
(\ref{milinv}).

\section{QFT on the open de Sitter universe \label{sec:dSQFT}}

\subsection{General considerations. Geometry.}
In this section we will apply our construction to the open de
Sitter universe. The simplest possible description of this
model is as follows: consider a $(d+1)$-dimensional Minkowski
spacetime with inner product
\begin{equation}
{X}\cdot {Y} = X^{0}{{Y}^{0}}-{X^{1}}{{Y}^{1}}-\ldots-
{X^{d}}{Y}^{d},
\end{equation}
and the embedded Lorentzian $d$-dimensional manifold $\Xd$ with
equation \begin{equation} \Xd= \{X: \,{X}^2 =  {X}\cdot  {X} =
-1\};\end{equation} this manifold models the whole de Sitter
universe (the grey manifold in Fig.   \ref{fig:opends}). Let us
consider now the intersection of $\Xd$ with the future cone
$V_+$ of a given event.  Specifically, we  consider the future
region of the event ${\O}=(0,\ldots0,1)$ (the "origin"):
\begin{equation} \Futcone = \Gamma^+_{ \rm O} =\{{\O} + V^+\}\cap \Xd =
\{X\in \Xd: \  (X-\O)^2 > 0,\ X^0>0\}.\end{equation} $\Futcone$
can be thought as a warped manifold, foliated with
$(d-1)$-dimensional Lobatchevskian leaves $\Lobad$. Precisely,
we have the following construction:
\begin{equation}
{X}({t},x) = \left\{\begin{array}{lcl}
X^{i} &=& \sinh {t} \,\,x^i ,\;\;i = 0,\ldots,d-1  \\
X^{d} &=&  \cosh {t}
\end{array}   \right.
\label{dScoorgenop}
\end{equation}
where $t>0$, and $x^2 = 1$, i.e. $x\in \Lobad$. In these
coordinates the de Sitter metric is written as follows:
 \begin{eqnarray}
 ds^2  &=&  \left. \left\{ {\left(dX^{0}  \right)}^2 -
{\left(dX^{1}  \right)}^2-\ldots -
{\left(dX^{{d}}  \right)}^2  \right\}  \right|_{\Xd} =  d{t}^2 - \sinh^2
{t} \, dl^2_{d-1};
 \end{eqnarray}
  this is a warped product with
warping function $\warp(t) = \sinh t$; in cosmology such a
metric defines a particular instance of a
Friedmann-Robertson-Walker hyperbolic universe. We will call
the region $\Futcone$ parametrized with the cosmic time given
chosen in (\ref{dScoorgenop})  the {\em open de Sitter
universe}  (Fig.   \ref{fig:opends}).
\begin{figure}[h]
\begin{center}
\includegraphics[height=6cm]{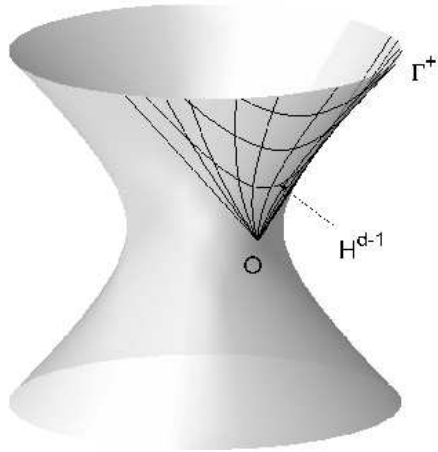}
\caption{\em The open de Sitter model is the interior of the future cone $\Gamma^+_{ \rm X}$
of any given event  on the de Sitter manifold (in the figure {X= \rm O}).
The surfaces of constant time are
 copies of the Lobatchevski space $\Lobad$. The geodesic worldlines of particles
 having constant spatial coordinates are  branches of hyperbolae.} \label{fig:opends}
\end{center}
\end{figure}

As briefly mentioned in the Introduction, the open de Sitter
universe was popular in the mid-nineties when it was playing a
central role in open models of inflation
\cite{Ratra94,kamionkowski,Bucher:1994gb}. These models were
abandoned when the microwave background fluctuation measurement
showed that our universe is most likely flat.  From both the
physical and mathematical viewpoints, correctly quantizing a
field in the open de Sitter manifold is an arduous task. In
particular, a naive application of the procedures of canonical
quantization gives a "wrong" result \cite{Sasaki,MS} in the
sense that when those procedures are applied to the open de
Sitter manifold one does not en up with the standard de Sitter
invariant vacuum \cite{Birrell,Gibbons,bgm,BM}. One reason is
that the spatial manifold $\Lobad$ represented in Fig.
  \ref{fig:opends} is a complete Cauchy surface for $\Futcone$
that is its own Cauchy development (i.e. its future), but it
fails to be so for the whole de Sitter manifold where it cannot
be used to set up the usual canonical quantization. In
\cite{Sasaki} this difficulty was circumvented by finding an
extension of the modes, originally defined only in the physical
region $\Futcone$, to the whole de Sitter manifold and applying
the canonical quantization there. The major drawback  of this
approach is that it is strictly limited to the de Sitter
geometry. The method used in \cite{MS} was not based on
canonical quantization but also necessitated the extension of
the open de Sitter manifold to the whole manifold.
 These calculations thus are flawed by the necessity, in order to explain local events,  to use (possibly complex) extentions of space-time to classically
unreachable regions, whose a priori  existence or non-existence in more general situation cannot be simply established.
This drawback is avoided in the present approach since by choice we work solely in the physical  space, here the open de Sitter manifold.

In the following we will perform in the open de Sitter universe
the same Fourier-type analysis already described in the Milne's
case. As before, an immediate bonus is the complete resolution
of the standard de Sitter QFT solely in terms of the modes of
the physical region $\Futcone$. One valuable aspect of the
method that we use is that it can also in principle be used to
analyze observational data. Our discussion is limited here to
the square-integrable case. Theories that involve modes that
are not square-integrable case as well as the implications of
our analysis for general canonical quantum theory will be the
object of separate studies.

\subsection{The complex de Sitter hyperboloid\label{sec:dScompl}}

In the study of quantum field theory on a given background, the
complexification of the underlying manifold plays a central
role in either the study of general properties (as for instance
the  PCT theorem) or in the construction of concrete models,
which are essentially based on Euclidean methods, i.e. on the
analytical continuation in the time coordinate. The de Sitter
case is no exception to this rule.

The complex de Sitter spacetime can be described as a complex
manifold embedded in the $(d+1)$-dimensional complex Minkowski
spacetime with equation:
\begin{equation} \Xdc =  \{Z  =  X+i Y \in {\Bbb M}^{(c)}:\,  Z\cdot Z = -1\} \label{hypec}
\end{equation}
As in the flat case (see Eqs.   \ref{MinkTmoins} and
  \ref{MinkTplus}) we introduce the tubular domains $\cal T^+$
and $\cal T^-$:
\begin{eqnarray} {\cal T}^{+}_d &=& \{ Z\in \Xdc: Y\cdot Y
> 0;\;\; Y^0 > 0\},\\
 {\cal T}^{-}_d  &=& \{ Z\in \Xdc: Y\cdot Y
> 0;\;\; Y^0 < 0\}, \label{tubi1} \end{eqnarray}
defined as the intersection of the complex Sitter manifold with
the the forward and backward tubes in the ambient complex
Minkowski spacetime. These domains arise in connection with the
thermal physical interpretation of de Sitter QFT \cite{BM,bem}.
More precisely, assumption of analyticity of the correlation
functions in (generalizations of) these domains give rise to
the KMS property and therefore to the thermal interpretation: a
de Sitter geodetic observer perceive a thermal bath of
"particles".

As in the Milne case, special attention will be devoted to the
complex events
 \begin{equation}
{Z}({\tau},x) = \left\{\begin{array}{lcl}
Z^{i} &=& \sinh\tau \,\,x^i ,\;\;i = 0,\ldots,d-1  \\
Z^{d} &=&  \cosh\tau
\end{array}   \right.
\label{Zdetau}
\end{equation}
where only the cosmic time has been complexified.  The complex
coordinate $\tau=t+is$ is defined in the strip
\begin{equation}
\Im\tau = s \in(-\pi,\pi)
\label{strip}
\end{equation}
of the complex $\tau$-plane. Events ${Z}({\tau},x)$ such that
$\Im\tau\in(0,\pi)$ belong to ${\cal T}^+$;  events
${Z}({\tau},x)$ such that $\Im\tau\in(-\pi,0)$ belong to ${\cal
T}^-$.

An alternative description of these events can be given by
using the variable $u = Z^d = \cosh \tau$. The image of both
$\cal T^+$ and $\cal T^-$ in the $u$-variable is the cut plane
\begin{equation}
\Delta = {\Bbb C}\setminus \{(-\infty,-1]\cup [1,\infty)\}
\label{cutplane}
\end{equation}
We are then led to consider the mappings $u \to {Z^\pm}({u},x)$
defined in $\Delta\times \Lobad$ as follows\footnote {The
change of complex variables $\tau\to u$ maps the strip
$\Im\tau\in(-\pi,\pi)$ onto a two sheeted manifold  defined by
the cuts of $\Delta$. With this mapping  we could have
considered a function $Z(u,x)$ that coincides with
${Z^+}({u},x)$ on a sheet and with ${Z^-}({u},x)$ on the other
sheet. Since the de Sitter global waves are defined either in
$\cal T^+$ or in $\cal T^-$  it is simpler to use the
(one-sheeted) cut-plane $\Delta$ and two different functions
mapping $\Delta$ to $\cal T^+$ and to $\cal T^-$. We use this
viewpoint throughout the present section.
\label{foot:utaumap}}:
\begin{equation}
{Z^\pm}({u},x) = \left\{\begin{array}{lcl}
Z^{l} &=& \pm i\left(1-u^2  \right)^\frac12 \,\,x^l ,\;\;l = 0,\ldots,d-1  \\
Z^{d} &=&  u
\end{array}   \right.
\label{Zplus}
\end{equation}
\begin{figure}[h]
\begin{center}
\includegraphics[height=5cm]{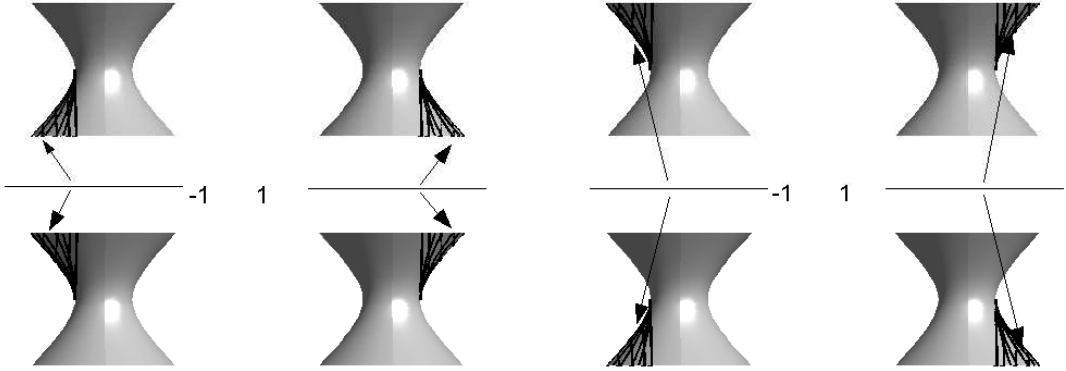}
\caption{\em  Two copies (left and   right of the figure) of the cut plane $\Delta$ of the
complex $u$ variable showing the cuts at $(-\infty, -1) { \rm U} (1,+\infty)$.
The 
images
of the de Sitter manifold correspond to  $u$
infinitesimally close to the cuts, at the place where the manifolds are drown. {\bf Left:}
 points of $\Delta$ are mapped through $Z^{-}({u},x)$  into ${\cal
T}^-$. The imaginary part of these points is  contained in $V^-$
(not represented). The copy at the lower   right is the
one which corresponds to the open de Sitter space that we solely
consider in the present paper. {\bf Right:}  points of $\Delta$
are mapped through $Z^{+}({u},x)$ into ${\cal T}^+$. The
imaginary part of these points is  contained in $V^+$ (not
represented). The copy at the upper  right is the one
which corresponds to the open de Sitter space that we solely consider
in the present paper. }
\end{center}
\end{figure}
It is readily seen  that $Z^{+}({u},x)\in {\cal T}^{+}$ and
$Z^{-}({u},x)\in {\cal T}^{-}$.  Indeed, consider for instance
the mapping $u \to {Z^+}({u},x)$. When $u$ is real and such
that $-1<u<1$, the events ${Z^+}({u},x)$ evidently belong to
the future tube ${\cal T}^{+}$ because for such events  $\Im
Z^+ \cdot \Im Z^+>0$ and  $\Im Z^0 = x^0\Im
\left(i\sqrt{1-u^2}  \right) >0$. Now, the first of these
conditions holds true for any $u\in\Delta$. On the other hand,
since the zeros of $\Im \left(i\sqrt{1-u^2}  \right)$ all belong
to the cuts of $\Delta$ one also has that
\begin{equation}
\Im Z^+({u,x}) = x^0 \Im \left(i\sqrt{1-u^2}  \right) >0 \;\;\makebox{for all} \;\; u\in\Delta,
\end{equation}
and the result follows. Note that we have ${Z^-}({u},x) =
\left[{Z^+}({u},x)  \right]{^*}$ (see App. (\ref{app:onthecut}),
Footnote (\ref{foot:star})).

\subsection{Plane waves.}

Let us consider an eigenfunction $\phi$ of the de Sitter
Klein-Gordon operator:
\begin{equation} \square \phi +m^2 \phi =0 \label{kgdS} \end{equation}
here $\square$ denotes the de Sitter-d'Alembert operator (i.e.
the Laplace-Beltrami operator relative to the de Sitter
metric). The usual approach to such an equation in curved
spacetimes consists in trying to solve it by separating the
variables in a suitably chosen coordinate systems; here in the
system (\ref{dScoorgenop}). If we do that and factorize the
wave $\phi$ by separating the variables according with the
reference system (\ref{dScoorgenop})
\begin{equation}
\phi(X) = f(t) \  \psi_{iq}(x,\xi) \label{factor}
\end{equation}
the time-dependent factor $f(t)$ is required to satisfy the
equation:
\begin{equation} \frac{1}{(\sinh t)^{d-1}}\frac{\d }{\d t} (\sinh t)^{d-1} \frac{\d f}{\d
t} + \frac{1}{(\sinh t)^2}\left[\left(\frac{d-2}{2}  \right)^2 +
q^2  \right] f + m^2 f =0.\label{tempcoord}
\end{equation}

There is also the possibility to introduce global waves in a
coordinate - independent way \cite{bgm,BM} by using the
embedding of the de Sitter hyperboloid in the Minkowski ambient
spacetime. Their construction is identical to that of the
spatial wavefunctions  (\ref{Sigwaves}) with an important
difference: they are singular on $(d-1)$-dimensional light-like
submanifolds of $\Xd$. This difficulty can be overcome by
moving to the complexification of the de Sitter spacetime
(\ref{hypec}).  The physically relevant global waves can be
defined as the functions
\begin{equation}
 \mbox{Const}  \left(Z \cdot { \Xi}   \right)^{-\frac{d-1}{2} + i\nu}
 \label{dSwaves}
 \end{equation}
where, as before, ${ \Xi}  = ( \Xi^{0},\ldots , \Xi^{d})$
belong to future lightcone in the ambieent spacetime, i.e. it
is a future directed null vector of the ambient space ($
\Xi\cdot \Xi = 0$ and $ \Xi^{0}>0$). The parameter $\nu$ is a
complex number. The physical values it may take are real, or
purely imaginary with $|\nu| \leq \frac{d-1}{2}$, and
correspond to
\begin{equation} m^2 =  \left(\frac{d-1}{2}  \right)^2 + \nu^2 \, \ge \, 0.
\label{mnurel} \end{equation}
The waves (\ref{dSwaves}) are analytic for $Z$ in the tubular
domains $ {\cal T}^{+}$ or $ {\cal T}^{-}$ of $\Xd^{(c)}$,
defined in (\ref{tubi1}). These analyticity properties are the
counterpart in the de Sitter universe of the {\em spectral
condition} of the Minkowski case \cite{BM,bem}.

\begin{figure}[h]
\begin{center}
\includegraphics[height=4cm]{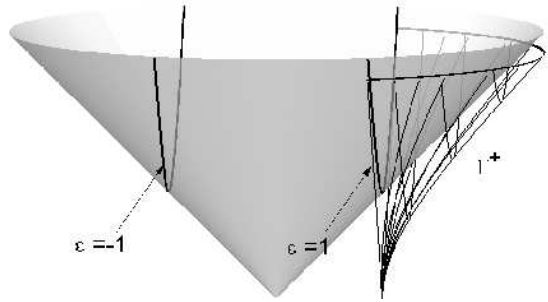}
\caption{\em The most convenient choice to represent the absolute of the open de Sitter space (shown at the  right) is by a two-component manifold,
labeled here by the two values of $\Xisign = \pm 1$.
 Each component is a copy of the Lobatchevski space $\Lobad$, similar to the mass shell of the Minkowski case.} \label{fig:opends3}
\end{center}
\end{figure}
%
%
Given the $(d+1)$-dimensional vector ${ \Xi}  = (
\Xi^{0},\ldots \Xi^{d})$ as above the $d$-dimensional vector
$(\Xi^0,\ldots,\Xi^{d-1})$  is timelike and forward directed.
One has that $(\Xi^0,\ldots,\Xi^{d-1}) = |\Xi^d|
(\Xiang^0,\ldots,\Xiang^{d-1})$ where $\Xiang^2 = 1$. There is
no loss of generality in setting $|\Xi^d|=1$; this is indeed
another possible choice to represent the absolute of the
ambient spacetime, and indeed the most convenient for our
purposes. This manifold has two disconnected components:
\begin{equation} \Xi \to (\Xiang, \Xisign), \  \Xi =
(\Xiang^0,\ldots,\Xiang^{d-1},  \Xisign ),\;\;\;\Xisign = \pm
1\;\label{parameter}\end{equation}
{From} (\ref{Zdetau}), we see that the scalar product
\begin{equation}
 Z(\tau,x)\cdot \Xi(\Xiang,\Xisign)=   \, x \cdot \Xiang\,\sinh {\tau} -
\Xisign   \, \cosh {\tau} , \label{sp2}
\end{equation}
has an imaginary part which does not vanish for events strictly
within ${\cal T}^+$ or within ${\cal T}^-$.
 The waves
(\ref{dSwaves}) are therefore globally (but separately)
well-defined in both
these domains. %

\subsubsection*{Representation of the plane waves.}

The  embedding of the de Sitter hyperboloid in the Minkowski
ambient spacetime is the foundation of the construction of the
global waves. The same embedding leads naturally
\cite{Bertola,MSfollow} to an integral representation of the de
Sitter waves (\ref{dSwaves}) in terms of the Minkowski
coordinates. With a convenient choice of phase and
normalization, the waves (\ref{dSwaves}) may be expressed in
the forward tube as follows:
\begin{eqnarray}
\left(-iZ\cdot\Xi  \right)^{-\frac{d-1}2 + i\nu} =
\frac{1}{\Gamma\left(\frac{{d-1}}{2}-i \nu  \right)}\, \int_{0}^{\infty} \frac{dR}{R} \
R^{\frac{d-1}{2} - i \nu} e^{iR\left(Z\cdot\Xi  \right)},\;\;\; Z\in {\cal T}^+.
\label{MelidSwavesT+}
\end{eqnarray}
The strictly positive imaginary part of $Z\cdot\Xi$ guarantees
the proper definition of the wave as well (see above) as well
as the convergence of the integral at infinity. The integral
converges also at the origin provided  $\vert\Im\nu\vert <
\frac{{d-1}}{2}$.

A similar representation holds in ${\cal T}^-$:
\begin{eqnarray}
\left(iZ\cdot\Xi  \right)^{-\frac{d-1}2 + i\nu} =
\frac{1}{\Gamma\left(\frac{{d-1}}{2}-i \nu  \right)}
\, \int_{0}^{\infty} \frac{dR}{R} \
R^{\frac{d-1}{2} - i \nu} e^{-iR\left(Z\cdot\Xi  \right)},\;\;\; Z\in {\cal T}^-.
\label{MelidSwavesT-}
\end{eqnarray}
Now it the strictly negative imaginary part of $Z\cdot\Xi$
that makes the integral converge.

\subsection{Analysis of plane waves in the open de Sitter universe. \label{sec:dSoscal}}

In the following we want to provide an expansion of de Sitter
waves in terms of the eigenmodes of the hyperbolic Laplacian.
As before the relevant expansion can be obtained by  by
computing the following integral transform
\begin{equation} F^{\pm}_{q,\nu}(\tau,\xi,\Xiang,\Xisign) = \int_{\Lobad}  d\mu(x) \  \left({x\cdot\xi}  \right)^{-{\frac{d-2}{2}} - iq}
\left(\mp i{Z(\tau,x)\cdot \Xi(\Xiang,\Xisign)}  \right)^{-\frac{d-1}2 +
i\nu}  . \label{dSosscalprod}
 \end{equation}
The integral is well-defined both for $0< \Im \tau < \pi$
(corresponding to $F^+$) and $-\pi< \Im \tau <0$ (corresponding
to $F^-$) so as to avoid the vanishing of $X(\tau,x)\cdot
\Xi(\Xiang,\Xisign)$. A glance to the large $x(r,\x)$ behavior
of (\ref{dSosscalprod}) that corresponds to $r\to0$, shows that
the simultaneous convergence of $F^{\pm}_{q,\nu}$ and
$F^{\pm}_{q,-\nu}$ is guaranteed by the condition
$\vert\Im\nu\vert < \frac12$, since we consider only real
values of $q$. This condition sets a lower bound of the masses
of the field theories that are covered by the present
treatment.

As before $SO_0(1,d-1)$ invariance  tells us  that result must
be function of the scalar product ${\xi\cdot \Xiang}$.
Homogeneity then  implies that
\begin{equation} F^{\pm}_{q,\nu}(\tau,\xi,\Xiang,\Xisign) =
f^{\pm}_{q,\nu}(\tau,\Xisign)\left({\xi\cdot \Xiang}  \right)^{-{\frac{d-2}{2}}
- iq}.
\end{equation}
The  functions $f^{\pm}_{q,\nu}(t,\Xisign)$ are therefore the
relevant solution of  equation (\ref{tempcoord}) that agree
with the spectral condition described above.

\subsection{Two-dimensional case.}

We start by discussing the two-dimensional case, which deserves
special consideration because of its simplicity. The spatial
manifold ${{\Bbb H}^1}$ is one-dimensional and can be
parametrized by an hyperbolic angle $v$:
\begin{equation}
x = (\cosh v, \sinh v),\;\;\;d\mu(x) = dv
\end{equation}
(i.e. $r = e^{-v}$ in Eq. (\ref{dscoor})). Labeling the modes
of the Laplacian is also quite simple: the spatial absolute has
only two possible direction; consequently, the spatial momentum
vector $\xi$ can take only two discrete values: $ \xi_l=
(1,-1)$ and $\xi_r=(1,1)$ so that $ \xi_l \cdot x = \cosh v +
\sinh v = e^v$ and $\xi_r\cdot x = e^{-v}$.

As regards the plane waves, they are labeled as in Eq.
(\ref{parameter}) by the discrete variable $\epsilon = \pm 1$
and by the one-dimensional vector $a$; the latter may in turn
be also parameterized by an hyperbolic angle $\Xiang= (\cosh w,
\sinh w)$ so that
\[x\cdot\Xiang= \cosh (v-w)\]
The computation of the scalar product is then straightforward
\begin{eqnarray} F^{\pm}_{q,\nu}({\tau},\xi_l,\Xiang,\Xisign)  &=&
 f^{\pm}_{q,\nu}({\tau},\Xisign) (\xi_l\cdot \Xiang)^{-iq}, \qquad F^{\pm}_{q,\nu}({\tau},\xi_r,\Xiang,\Xisign)  =
f^{\pm}_{q,\nu}({\tau},\Xisign) (\xi_r\cdot \Xiang)^{-iq},
\cr  \cr f^{\pm}_{q,\nu}({\tau},\Xisign)
&=&\int_{-\infty}^{+\infty} dw \ e^{-iqw}
\left(\mp i\cosh w\ \sinh {\tau} \pm i \Xisign    \,\cosh {\tau}  \right)^{-\frac{1}2+ i\nu}.
 \label{FdS2}
\end{eqnarray}
The two-dimensional case also provides  an easy direct check of
the inversion formula. Indeed, with the above choice of
coordinates, it simply amounts to Fourier inversion.

\paragraph*{Expression of $f$ in terms of Legendre functions.}

The functions $f^{\pm}_{q,\nu}({\tau},\Xisign)$ are completely
characterized by the integral representation (\ref{FdS2}). They
can be however expressed in terms of the associated Legendre
functions. We use here and in the following the notations and
the conventions of the Bateman manuscript project \cite{BAT}
with one notable exception (the function $\Q$, see below). In
particular the Legendre functions $P$ and $Q$ are assumed to be
analytic and one valued on the complex plane cut from
$z=-\infty$ to $z=1$. The two cases to be considered look
however at first rather different, since the integrand never
vanishes for $\Xisign=-1$, while  it become singular along the
integration path  for $\Xisign=1$ (for $\tau$ real: $\tau=t
>0$).

\paragraph*{Case $\Xisign=-1$.}

In this case the integral representation (\ref{FdS2}) already
coincides with a well-known integral representation of a
Legendre functions of the second kind \cite{BAT}  Eq.
(3.7;12):
\begin{eqnarray} f^{\pm}_{q,\nu}({t},\Xisign=-1)  =
 2e^{\pm i\frac\pi2(\frac12-i\nu)}    \frac{\Gamma\left(\frac12 -
i\nu-iq  \right)}{\Gamma\left(\frac12 - i\nu  \right)}
e^{\pi q} Q^{iq}_{-\frac{1}{2}-i\nu}(\cosh t)
\label{fdS2near}
\end{eqnarray}

\paragraph*{Case $\Xisign=1$.} Here the integral in (\ref{FdS2}) may be
split into two parts according to the sign of the expression
$\cosh w\ \sinh {\tau} - \cosh {\tau}$: one addendum  may be
evaluated by means of \cite{BAT} Eqs. (3.7;8) and  (3.3;14)
and the second by means of \cite{BAT} Eqs. (3.7;5) and
(3.3;13), to yield
\begin{eqnarray} \lefteqn{f^{\pm}_{q,\nu}({\tau}, \Xisign=1)
= \frac{2i e^{\pm i\frac\pi2(\frac12-i\nu)} } { \sinh \pi \nu }
\frac{\Gamma\left(\frac12 - i\nu-iq  \right)}{\Gamma\left(\frac12 -
i\nu  \right)}} \cr && \times \left[ e^{\mp \pi\nu } {\cosh\left(\pi
\left(q+\nu  \right)  \right)}\ e^{\pi q} Q_{-\frac{1}{2}+i \nu }^{i q}(\cosh t)
-{\cosh \pi q }\ e^{\pi q} Q_{-\frac{1}{2}-i \nu }^{i q}(\cosh
t)  \right]
\label{fdS2far}
\end{eqnarray}

\paragraph*{Expression of $f$ in terms of Legendre functions "on the
cut". \label{sect:f2ocut}}

There is a more elegant way and synthetic way to express the
modes in terms of Legendre functions  based on the use of the
variable $u = Z^d$ introduced in Sect.   \ref{sec:dScompl}}. This
alternative procedure  has also the advantage to fully exhibit
the underlying symmetries.  The input of the relations
(\ref{Zplus}) of $Z$ into  the expressions of the de Sitter
waves (\ref{MelidSwavesT+}) and (\ref{MelidSwavesT-}) gives:
\begin{equation}
f^\pm_{q,\nu}(u,\Xisign) =
\int_{-\infty}^{\infty}dv \  e^{- i qv}\left[\left(1-u^2  \right)^\frac12\cosh v   \pm \Xisign i u
  \right]^{-\frac{1}{2}+i \nu} . \label{scalint}
\end{equation}
The two functions $f^\pm_{q,\nu}(u,\Xisign) $ are manifestly
analytical in the cut plane $\Delta$ introduced in Eq.
(\ref{cutplane}): the term in square brackets at the RHS
vanishes for $u = \coth v$ for real $v$, and therefore the
integral is well-defined for $u \notin (-\infty,-1)\cup
(1,+\infty)$ with no additional singularity in $\Delta$. We see
once more that the domain $\Delta$ is naturally related to the
tuboids of analyticity of the de Sitter waves (\ref{dSwaves}).
It is therefore natural to make use of the following "Legendre
function on the cut" (see App.   \ref{app:onthecut}):
\begin{eqnarray}
\P^{iq}_{-\frac{1}{2}-i\nu} (u) &=& e^{\mp \frac{1}{2} \pi q }
P^{iq}_{-\frac{1}{2}-i\nu} (u), \label{Pcut}
\\ \Q^{iq}_{-\frac 12 - i\nu} ( u) &=&   \frac{e^{ \frac{i\pi}2(-\frac12+i\nu)}}{2\sin i\pi{q}  }
 \left[ \frac{e^{-\frac12 \pi q}\ \P^{iq}_{-\frac 12 + i\nu} ( u)}{\Gamma\left(\frac 12 - i\nu+iq  \right)} -
\frac{e^{\frac12 \pi q}\  \P^{-{iq}}_{-\frac 12 + i\nu}( u)}{\Gamma\left(\frac 12 -i\nu-{iq}  \right)}   \right]
 . \label{Qcut}
 \end{eqnarray}
The upper or lower signs of (\ref{Pcut}) refer to the imaginary
part of $u$ being positive or negative. These functions are
analytic in the cut plane $\Delta$. A brief summary of their
properties and symmetries is given in Appendix
(\ref{app:Qfunct}). We then see, with $\Q{{^*}}^{iq}_{-\frac
12 + i\nu} (u)  = \Q^{iq}_{-\frac 12 - i\nu} (- u) $, that Eqs.
(\ref{fdS2near}) and (\ref{fdS2far}) are conveniently expressed
as:
\begin{eqnarray}
f^+_{q,\nu}(u,\Xisign) &=&
 \frac{{2\pi\Gamma\left(\frac 12 - i\nu+iq  \right)}{\Gamma\left(\frac 12 -i\nu-{iq}  \right)}}
{\Gamma\left(\frac{1}{2}-i \nu  \right)} \  \Q^{iq}_{-\frac 12 - i\nu} (\Xisign u),
\;\;\;\;\;\;\; \cr
 f^-_{q,\nu}(u,\Xisign) &= & \frac{{2\pi\Gamma\left(\frac 12 - i\nu+iq  \right)}{\Gamma\left(\frac 12 -i\nu-{iq}  \right)}}
{\Gamma\left(\frac{1}{2}-i \nu  \right)} \  \Q{{^*}}^{iq}_{-\frac 12 + i\nu} (\Xisign u) .
\label{fdS2uQ}
\end{eqnarray}

\subsection{Any dimension}
\label{sec:dSscalint} Now that the two-dimensional case has
been solved the general $d$-dimensional can be faced more
easily. Let us go back to the complex time variable $\tau$ and
consider say $\Im \tau
>0$. By using the parametrization (\ref{dscoor}), the scalar
product $X \cdot \Xi$ and the integral representation
(\ref{MelidSwavesT+}) may be written as follows:
\begin{equation}
{x( r,\x) \cdot \Xiang( \rho, \aa)} \sinh \tau - {\Xisign} \cosh \tau =
\frac{1}{{ \rho}}\left[\frac{({\x} - {\aa})^2 + r^2 +  \rho^2}{2r} \sinh
\tau - \Xisign  \rho\cosh \tau  \right], \label{explicit1}
\end{equation}
\begin{eqnarray}
&&\left(-iZ\cdot\Xi  \right)^{-\frac{d-1}2 + i\nu} = \left( -i x \cdot \Xiang \, \sinh \tau+i\Xisign  \, \cosh
\tau  \right)^{-\frac{d-1}2 +i\nu} = \cr &&=  \frac{ \rho^{{\frac{d-1}2
-i\nu}}}{\Gamma\left(\frac{{d-1}}{2}-i \nu  \right)}
 \, \int_{0}^{\infty} \frac{dR}{R} \
 R^{\frac{d-1}{2} - i \nu} e^{iR\,\left[\frac{(\x - \aa)^2 +  r^2 +  \rho^2}{2 r}
\sinh \tau -\Xisign  \rho\cosh \tau  \right]}.
 \label{MelidSwavesbis}
\end{eqnarray}
Given this formula, the steps to compute
$F^{+}_{q,\nu}(\tau,\xi,\Xiang,\Xisign)$ are similar to those
of the Milne case.  At first the integral representations
(\ref{comprep}) and (\ref{MelidSwavesbis}) are inserted into
Eq. (\ref{dSosscalprod}) and perform the Gaussian integral over
$\x$. By using the same change of variables as in (\ref{path})
and (\ref{path2}) and identifying one factor with (the integral
representation of) $\left({ \Xiang \cdot\xi}  \right)^{-{{\frac{d-2}{2}}} -
iq}$ gives
\begin{eqnarray}
 F^{+}_{q,\nu}(\tau,\xi,\Xiang,\Xisign)  &=& \frac{\Gamma\left(\frac{1}{2}-i \nu  \right)}
{\Gamma\left(\frac{d-1}{2}-i \nu  \right)}
 \left(\frac{2\pi i }{\sinh\tau}  \right)^{{\frac{d-2}{2}}}
\left({\Xiang \cdot\xi}  \right)^{-{{\frac{d-2}{2}}} - iq} \times \cr &&
\times \int_{0}^{\infty} \frac{dR}{R} \  R^{- i q}\left[-i\left(
\frac{R}{2}+\frac{1}{2R}  \right)\sinh \tau+i\Xisign \cosh \tau
  \right]^{-\frac{1}{2} + i \nu} . \label{scalint+}
\end{eqnarray}
As before, the integral at the RHS is symmetric in the exchange
$q\to-q$. This result holds for  $0<\Im \tau<\pi$.  The other
case  $-\pi<\Im\tau<0$ can be obtained similarly. It follows
that
\begin{eqnarray}
f^{\pm}_{q,\nu}(\tau,\Xisign) &=&
\frac{\Gamma\left(\frac{1}{2}- i \nu  \right)} {\Gamma\left(\frac{d-1}{2}- i \nu  \right)}
 \left(\frac{\pm2\pi i}{\sinh\tau}  \right)^{{\frac{d-2}{2}}} \times \cr &&
\times
 \int_{0}^{\infty} \frac{dR}{R} \  R^{- i q}\left[
\mp i \left(
\frac{R}{2}+\frac{1}{2R}  \right)\sinh \tau \pm i \Xisign \cosh \tau   \right]^{-\frac{1}{2} + i \nu}
\label{osdSscal'}
 \end{eqnarray}
 where $f^+$
is defined in the strip $0< \Im \tau <\pi$ while $f^-$ is
defined in the strip $-\pi <\Im \tau <0$.

The inversion formula  gives the expansion that we have worked
so hard to obtain:
\begin{equation}
 \left(\mp i Z  \cdot { \Xi}   \right)^{-\frac{d-1}{2} + i\nu} =
 \int_{0}^{\infty}  {\Nq(q)}{dq }\   \int d\mu(\xi) \
 f^{\pm}_{q,\nu}(\tau,\Xisign) \left(\Xiang \cdot\xi  \right)^{-{\frac{d-2}{2}} - iq}
 \left({x \cdot \xi}   \right)^{- \frac{d-2}{2} +i  {q}}
 \label{dSwaveexp}
\end{equation}
 The upper or lower sign is to be taken accordingly as $\Im \tau>0$ or $\Im \tau <0$.
As in the Milne case (Sect.   \ref{sec:Milinv}), it is more
concise to rewrite this expansion as a one-dimensional integral
by means  of Eq. (\ref{projq}) the projector $\PR_q(\Xiang,x)$
onto the space of open waves with eigenvalue $q^2$. The latter
is a particular solution of the Laplace equation, which depends
on the de Sitter wave parametrisation since it is labeled  by
$\Xiang$.
 \begin{equation}
 \left(\mp i Z  \cdot { \Xi}   \right)^{-\frac{d-1}{2} + i\nu}
= \int_0^\infty  dq \ f^{\pm}_{q,\nu}(\tau,\Xisign)  \PR_q(\Xiang,x) .
\label{dSosexp}
 \end{equation}
Note the analogy with the Milne case: here the role of $p/m$ is
played by the parameter $\Xiang$.

The whole discussion of the two-dimensional case can be
repeated and we get expressions for $ f^\pm_{q,\nu}(u,\Xisign)$
in terms of the Legendre functions "on the cut":
\begin{eqnarray}
&& f^+_{q,\nu}(u,\Xisign) = \left({2\pi}  \right)^\frac d2
 \frac{{\Gamma\left(\frac 12 - i\nu+iq  \right)}{\Gamma\left(\frac 12 -i\nu-{iq}  \right)}}
{\Gamma\left(\frac{d-1}{2}-i \nu  \right)}  \left(1-u^2  \right)^{-\frac{d-2}{4}}\Q^{iq}_{-\frac 12 - i\nu} (\Xisign u)
\cr
 &&f^-_{q,\nu}(u,\Xisign) = \left({2\pi}  \right)^\frac d2
 \frac{{\Gamma\left(\frac 12 - i\nu+iq  \right)}{\Gamma\left(\frac 12 -i\nu-{iq}  \right)}}
{\Gamma\left(\frac{d-1}{2}-i \nu  \right)}  \left(1-u^2  \right)^{-\frac{d-2}{4}}\Q{{^*}}^{iq}_{-\frac 12 + i\nu} (\Xisign u)
 \,\,\,\,\,\,\,\,\,
\label{fdSuQ} \cr
&&
 \end{eqnarray}
that are  analytical functions defined on the cut plane
$\Delta$.

\subsection{Expansion of the de Sitter Wightman function}

The more elegant way to write the de Sitter two-point Wightman
function of a massive Klein-Gordon field is as a superposition
of the global waves (\ref{dSwaves}) \cite{BM}:
\begin{equation} W(Z,Z') = \cnew_{d,\nu} \int d\mu(\Xi) \
\left(iZ \cdot { \Xi}   \right)^{-\frac{d-1}{2} + i\nu}
\left(-iZ'  \cdot { \Xi}   \right)^{-\frac{d-1}{2} - i\nu}
\label{dSWf}\end{equation}
\begin{equation} \cnew_{d,\nu} = \frac {
{{\Gamma\left(\frac{d-1}{2}+i\nu  \right)}{\Gamma\left({\frac{d-1}{2}}-i\nu  \right)}}}
{{2(2\pi)^{ d}}}. \end{equation} where  $Z \in T^-$ and
 $Z' \in T^+$. This writing is the one that is most similar to
 the Fourier plane wave expansion of the two-point function of
 the flat case (\ref{01}).

By inserting in this representation formulae (\ref{dSosexp})
one gets the spectral density $ \rho$ that provide the expansion
of the Wightman function in terms of the modes of the
Lobatchevskian Laplace-Beltrami operator:
\begin{equation}
W =  \int_{0}^\infty dq \
 \rho(q,\cosh \tau,\cosh \tau') \ \PR_q(x,x')  .
\end{equation}
with
\begin{eqnarray}
 \rho(q,u,u') &=&  \frac {{{\Gamma\left(\frac 12 - i\nu-iq  \right)}{\Gamma\left(\frac 12 + i\nu-iq  \right)}{\Gamma\left(\frac 12 - i\nu+iq  \right)}{\Gamma\left(\frac 12 + i\nu+iq  \right)}}}
{ 2 \  \left(1-u^2  \right)^{-\frac{d-2}{4}} \left(1-u'^2  \right)^{-\frac{d-2}{4}} }
 \cr  &\times&  \left[
 \Q^{iq}_{-\frac 12 - i\nu}(u) \Q{{^*}}^{iq}_{-\frac 12 - i\nu} (u') + \Q^{iq}_{-\frac 12 - i\nu}(-u) \Q{{^*}}^{iq}_{-\frac 12 - i\nu} ( - u')
  \right]. \label{dSSpec1}
\end{eqnarray}
The spectral density $ \rho(q,u,u')$ may also be expressed in
terms of the functions $\P^{iq}_{-\frac 12 + i\nu}(u)$ (see
Appendix (\ref{app:Qfunct})) as follows:
\begin{eqnarray}
  \rho(q,u,u') = && \frac {{{\Gamma\left(\frac 12 - i\nu-iq  \right)}{\Gamma\left(\frac 12 + i\nu-iq  \right)}{\Gamma\left(\frac 12 - i\nu+iq  \right)}{\Gamma\left(\frac 12 + i\nu+iq  \right)}}}
{4\pi \  \left(1-u^2  \right)^{-\frac{d-2}{4}} \left(1-u'^2  \right)^{-\frac{d-2}{4}}}
 \cr &&\times  \left[
 \P^{iq}_{-\frac 12 + i\nu}(u)\P^{-iq}_{-\frac 12 + i\nu}(u') + \P^{iq}_{-\frac 12 + i\nu}(-u)\P^{-iq}_{-\frac 12 + i\nu}(-u')
  \right]. \label{dSSpec2}
  \end{eqnarray}
These formulae provide the full solution of the difficult
problem of describing the quantum Klein-Gordon field on the
open de Sitter universe, provided that $\vert\Im\nu\vert <
\frac{1}2$ (see Sect.   \ref{sec:dSoscal}). The latter
requirement can be expressed as a condition on the mass
parameter according to (\ref{mnurel}):
\begin{equation}
m>m_{cr} =\frac{d(d-2)}{4} .
\end{equation}
This is the condition that guarantees the convergence of the
integral (\ref{dSosscalprod}). When $m<m_{cr}$  modes that are
not square-integrable are necessary \cite{Sasaki,MS,MSfollow} for a full
and correct description.

\section{Conclusions}

In this paper we have settled down the basic ingredients to
work out quantum theories on homogeneous and isotropic spaces
with negative curvature.

 {}{While the state-of-the-art may be considered to be the expansion
of the eigenmodes of the Laplacian in terms of spherical harmonics (see e.g. \cite{Ratra94}), we give here a new set of eigenmodes, based on a different decomposition. Most important, we provide also the way to deal with these new modes, that we believe is simpler than the standard approach. Simplicity is very often the source of new progress, that however we have left for future work.}

 {}The formalism we have developped here} is the closest possible
to the one employed in standard (textbook) quantum mechanics on
the Euclidean space ${\Bbb R}^3$, and is based on the suitable
Fourier-type harmonic analysis in terms of the eigenmodes of
the Laplacian precisely in the same way as standard quantum
mechanics is based on the Fourier transform in terms of plane
waves, i.e. the momentum space representation of the
wavefunctions. However, even though we have restricted our
attention to square-integrable functions, the absence of
translation invariance renders the task considerably more
complicated than in flat space. The eigenmodes of the Laplacian
are labeled in the most convenient way by using \cite{gelfand}
vectors of the cone asymptotic to the Lobatchevski space and a
real number $q$. An important technical point consist in
finding suitable integral representations expressing such
family of eigenmodes. This step is in the spirit of our earlier
work \cite{Bertola}, where general embedded manifolds (branes)
were studied. These integral representations render quite
simple calculations which at first sight might seem
intractable.

In the second part of this paper we have applied our general
construction to the study of two geometries which are  quite
popular nowadays: the universes of Milne and de Sitter. In both
cases we have displayed explicit formulae yielding the
Fourier-type expansion of the relevant spacetime plane waves in
terms of the eigenmodes of the Lobatchevski Laplacian. These
expansions immediately provide also harmonic expansions for the
corresponding Wightman functions in terms of representations of
the principal series of the $SO_0(1,d-1)$ group. One important
point of our treatment of the de Sitter case is that the result
is obtained  by working solely in the physical region covered
by the open chart,  in contrast with previous approaches
\cite{Sasaki,MS} that achieved a similar goal. The apparent
simplicity of our approach should not send to oblivion the fact
the whole construction was thought impracticable and matter of
big controversies.

The quantization of theories implying the use  of non-square
integrable modes remains to be attacked. A clean discussion of
the square-integrable case as presented here is a necessary
step forward. We expect that our methods and results will allow to proceed one step further and to tackle the analysis of a larger class of non square integrable functions.

 Due to their simplicity, the methods outlined in this paper set the
premises of many future developments in various other, quite
different, directions.

\newpage

\appendix

\section{Legendre functions "on the cut" \label{app:onthecut}}

\subsection{Legendre functions of the first kind}

Let us introduce the function
\begin{equation}
\P^{iq}_{-\frac{1}{2}+i\nu} (u)  =  e^{\mp \frac{1}{2} \pi q }
P^{iq}_{-\frac{1}{2}+i\nu} (u), \label{Pcutapp}
\end{equation}
where the upper or lower signs refer to the imaginary part of
$u$ being positive or negative. The function $\P$  is an
analytic continuation of the so-called "Legendre function on
the cut", originally defined for real $u$ such that $|u|<1$
\cite{BAT}, to the whole cut-plane $\Delta$ introduced in
(\ref{cutplane}). $\P^{iq}_{-\frac{1}{2}+i\nu} (u)$  is an
entire function of the complex parameters $q$ and $\nu$ (the
only singularities being at infinity). The functions
$\P_{-\frac 12 + i\nu}^{iq}(-u)$ and
$\P{{^*}}^{iq}_{-\frac{1}{2}+i\nu}(u) =
\P^{-iq}_{-\frac{1}{2}+i\nu}(u) $ are two other
solutions\footnote{We define $\chi{^*}(u) =
\overline{\chi(\overline{u})}$ where the bar denotes complex
conjugation. \label{foot:star}}
of the Legendre equation analytic on the same domain $\Delta$;
one has the following relation:
\begin{eqnarray}
\frac{1}{\Gamma\left({iq}  \right)\Gamma\left(1-iq  \right)}   \P^{iq}_{-\frac{1}{2}+i\nu}(-u)
 &=&
\frac{1}{\Gamma\left(\frac{1}{2}+i\nu  \right)\Gamma\left(\frac{1}{2}-i\nu  \right)}\P^{iq}_{-\frac{1}{2}+i\nu}(u) - \cr &&
\frac{1}{{\Gamma\left(\frac{1}{2}-i\nu-iq  \right)\Gamma\left(\frac{1}{2}+i\nu -i q  \right)}} \P^{-iq}_{-\frac{1}{2}+i\nu}(u)
 .  \label{Pcmq}
\end{eqnarray}

The following formula is useful to derive the spectral density
of the de Sitter case and can be obtained from the relation
(\ref{Pcmq}):
\begin{equation}
\frac {\P^{iq}_{-\frac{1}{2}+i\nu}(u)
\P^{iq}_{-\frac{1}{2}+i\nu}(-u')}
{\Gamma\left(\frac{1}{2}+{iq}+i\nu  \right)\Gamma\left(\frac{1}{2}+{iq}-i\nu  \right)}  =
\frac
{\P^{iq}_{-\frac{1}{2}+i\nu}(u)\P^{-iq}_{-\frac{1}{2}+i\nu}(-u') }
{\Gamma\left(\frac{1}{2}+i\nu  \right)\Gamma\left(\frac{1}{2}-i\nu  \right)} + \frac
{\P^{iq}_{-\frac{1}{2}+i\nu}(u)\P^{-iq}_{-\frac{1}{2}+i\nu}(u') }
{\Gamma\left({iq}  \right)\Gamma\left(1-{iq}  \right)},
\label{PP1}
\end{equation}
or
\begin{equation}
\frac {\P^{-iq}_{-\frac{1}{2}+i\nu}(u)
\P^{-iq}_{-\frac{1}{2}+i\nu}(-u')}
{\Gamma\left(\frac{1}{2}-{iq}+i\nu  \right)\Gamma\left(\frac{1}{2}-{iq}-i\nu  \right)}  =
\frac {\P^{iq}_{-\frac{1}{2}+i\nu}(u)
\P^{-iq}_{-\frac{1}{2}+i\nu}(-u')}
{\Gamma\left(\frac{1}{2}+i\nu  \right)\Gamma\left(\frac{1}{2}-i\nu  \right)} - \frac
{\P^{iq}_{-\frac{1}{2}+i\nu}(-u)\P^{-iq}_{-\frac{1}{2}+i\nu}(-u') }
{\Gamma\left({iq}  \right)\Gamma\left(1-{iq}  \right)}.
\label{PP2}
\end{equation}

\subsection{Legendre functions of the second kind
\label{app:Qfunct}}

Let us  define the  function $\Q^{iq}_{-\frac 12 - i\nu} ( u)$
by the following integral representation:
\begin{eqnarray}
\Q^{iq}_{-\frac 12 - i\nu} ( u) =
\frac{\Gamma\left(\frac{1}{2}-i \nu  \right)}{2\pi{\Gamma\left(\frac 12 - i\nu+iq  \right)}{\Gamma\left(\frac 12 -i\nu-{iq}  \right)}}
\ \int_{-\infty}^{\infty}dv \  e^{- i qv}\left[\left(1-u^2  \right)^\frac12\cosh v  + i u
  \right]^{-\frac{1}{2}+i \nu}. \cr \label{Qdef}
\end{eqnarray}
The function $\Q^{iq}_{-\frac 12 - i\nu} (u)$ is readily seen to be a solution of the Legendre equation. It is manifestly analytic the
cut-plane $\Delta$ and invariant under the change $q\to-q$. It
is proportional to the standard Legendre function
$Q^{iq}_{-\frac 12 + i\nu} ( u)$ \cite{BAT} in the upper $u$
half-plane:
\begin{equation}
\Q^{iq}_{-\frac 12 - i\nu} ( u) = \frac {e^{-
i\frac\pi2(\frac12-i\nu)}  }  {\pi{\Gamma\left(\frac 12 - i\nu+iq  \right)}}
e^{\pi q} Q^{iq}_{-\frac{1}{2}-i\nu}(u), \qquad \Im u
> 0;
\end{equation}
note however that the restriction of $\Q^{iq}_{-\frac 12 -i\nu}
( u)$ to real $u$ such that $|u|<1$ does not coincide with  the
"Legendre function on the cut" of the second kind as defined in
\cite{BAT}. Obviously, $\Q$ may be expressed in terms of any two
independent functions of the first kind and indeed, by using
\cite{BAT} Eq. (3.7;24), there follows that
\begin{eqnarray}
\Q^{iq}_{-\frac 12- i\nu} ( \pm u) &=&
  \frac{e^{\mp i \frac\pi2(\frac12-i\nu)}}{2\sin i\pi{q}  }
 \left[ \frac{e^{\mp\frac12 \pi q}}{\Gamma\left(\frac 12 - i\nu+iq  \right)} \P^{iq}_{-\frac 12 + i\nu} ( u)-
\frac{e^{\pm\frac12 \pi q} }{\Gamma\left(\frac 12 -i\nu-{iq}  \right)}\P^{-{iq}}_{-\frac 12 + i\nu}( u)   \right] =
\cr& =&
  \frac{e^{\pm i \frac\pi2(\frac12-i\nu)}}{2\sin i\pi{q}  }
 \left[ \frac{e^{\pm\frac12 \pi q}}{\Gamma\left(\frac 12 - i\nu+iq  \right)} \P^{iq}_{-\frac 12 + i\nu} ( -u)-
\frac{e^{\mp\frac12 \pi q} }{\Gamma\left(\frac 12 -i\nu-{iq}  \right)}\P^{-{iq}}_{-\frac 12 + i\nu}( -u)   \right]
 \label{QQstar}\cr &&
\end{eqnarray}

\end{document}